\documentclass[aps,prd,preprint,showpacs,superscriptaddress]{revtex4}
\usepackage{graphics}
\usepackage{graphicx}
\usepackage{epsfig}
\usepackage{amsmath}
\usepackage{amssymb}
\usepackage{amsthm}
\usepackage{calc}
\newcommand{\beq}{\begin{equation}}
\newcommand{\eeq}{\end{equation}}
\newcommand{\bea}{\begin{eqnarray}}
\newcommand{\eea}{\end{eqnarray}}
\newcommand{\tinyfrac}[2]{\frac{\text{\tiny#1 }}{\text{ \tiny #2}}}
\newtheorem{prop}{Proposition}

\begin{document}
\title{Entanglement classification of three fermions with up to nine single-particle states}
\author{G\'abor S\'arosi and P\'eter L\'evay}
\affiliation{Department of Theoretical Physics, Institute of
Physics, Budapest University of Technology and Economics, H-1521 Budapest,
Hungary}
\date{\today}
\begin{abstract}
Based on results well known in the mathematics literature
but have not made their debut to the physics literature yet we
conduct a study on three-fermionic systems with six, seven, eight
and nine single-particle states. Via introducing special
polynomial invariants playing the role of entanglement measures
the structure of the SLOCC entanglement classes is investigated.
The SLOCC classes of the six- and seven-dimensional cases can elegantly be described
by special subconfigurations of the Fano plane.
Some special embedded systems containing distinguishable
constituents are  arising naturally in our formalism, namely,
three-qubits and three-qutrits. In particular the three fundamental invariants $I_6$, $I_9$, and $I_{12}$ of the three-qutrits system are shown to arise as special cases of the four fundamental invariants of three-fermions with nine single-particle states.
\end{abstract}
\pacs{ 03.67.-a, 03.65.Ud, 03.65.Ta, 02.40.-k} \maketitle{}

\tableofcontents

\section{Introduction}
Quantum entanglement is a key resource for implementing tasks for processing quantum information\cite{Nielsen}.
It is well-known by now that this resource can be based on manipulating composite quantum systems with both distinguishable and indistinguishable constituents.
Though historically the study of entanglement based on systems belonging to the former class has received much greater scrutiny, investigations focussing on the latter have gained considerable attention too\cite{Sch1,Sch2,Sch3,You,Li,Ghir}. Quite recently fermionic
systems started to play a key role in studies
revisiting the so called N representability\cite{Coleman}
and quantum marginal problem\cite{Borland} centered around
studies employing the important notion of
entanglement polytopes\cite{Christandl,Christandl2,Sawicki}
an idea having roots in the work of Klyachko\cite{Kly1}.
The introduction of this notion was partly motivated by the study of special tripartite fermionic systems having
six, seven and eight single particle states\cite{Borland}.
These systems provide simple special examples for multifermionic wavefunctions with physical properties easy to investigate. On the other hand they also
give rise to mathematical structures, namely three-forms in a six, seven, eight
and nine
dimensional vector spaces over a field, well-known to mathematicians\cite{Reichel,Schouten,Gurevich1,Gurevich2,Cohen,Westwick,Dokovic,Vinberg}.
Though the results in these papers on the classification of trivectors bears a relevance on the so called SLOCC classification of entanglement classes\cite{Bennett,Dur} in quantum information, apart from scattered remarks\cite{Kly4,Djok2}  and our recent paper on Hitchin functionals\cite{levsar} to our best knowledge these systems have not made their full debut to the
literature on quantum entanglement.

The aim of the present paper is to present a study on these special entangled fermionic systems based on these findings.
In quantum information one wishes to quantify and classify different types of entanglement regarded as a resource.
There are different classification schemes. In the SLOCC classification scheme of multipartite systems the representative {\it pure} states are equivalent if they can be mutually converted to each other with a finite probability of succes using only local operations and classical commmunication.
It can be shown\cite{Dur} that for a system consisiting of $n$ distinguishable subsystems SLOCC equivalence  mathematically means that
the equivalent pure states representing the system are on the same orbit under the action of the local group $GL(N_1)\times GL(N_2)\times \cdots \times GL(N_n)$, where $(N_1,N_2,\dots N_n)$ are the local dimensions of the Hilbert spaces associated to the subsystems.
For systems with indistinguishable constituents the correponding orbit should be formed under the $n$-fold {\it diagonal} action of $GL(N)$ where $N$ is the number of single particle states.
Although due to proliferation of entanglement classes solving the SLOCC classification problem in its full generality
is a hopeless task, we still have a number of important special cases for which the structure of the SLOCC classes is known.
These special entangled systems can provide a convenient theoretical
framework to see multipartite entanglement in action.

Now although these special systems have already been studied by mathematicians however,
physicists are either not aware of these results or they are reluctant to apply them, or they are rediscovering them from time to time in different contexts.
For example the classification problem equivalent to the classification of SLOCC
 entanglement types for three qubits has already been solved in 1881 by mathematicians\cite{Lepaige} (see also the paper of Schwartz\cite{Schwartz} and the book of Gelfand\cite{Gelfand}), the result has later been independently rediscovered in the influencial paper by physicists\cite{Dur}.
As another example one can consider the case of three fermions with six single particle states a system used by Borland and Dennis in their seminal paper\cite{Borland}. Using results known from cubic Jordan algebras the SLOCC classes for this case has been rediscovered by one of us\cite{levvran1}. We have learnt
later that the solution to this problem dates back as early as 1907 via the work of Reichel. Moreover it also turns out that this case is also well-known from the theory of prehomogeneous vector spaces\cite{Kimura} where in the full classification of these spaces as given by Sato and Kimura
this type of fermionic systems shows up as an important special case\cite{Satokimura}.
As we already mentioned this case also constituted the archetypical example for further studies on entanglement polytopes and the N-representability problem
\cite{Christandl,Christandl2,Sawicki}.
Moreover, elevating a {\it real} three fermionic state with six single particle states to a three-form living on a six dimensional manifold renders the square root of the magnitude of the quartic
entanglement measure\cite{levvran1}  to a functional on the manifold\cite{levsar}.
As shown by Hitchin\cite{stable,Hitchin} in an important special case the critical points of this functional correspond to Calabi-Yau manifolds.
On the other hand the evaluation of this functional at the critical point gives the semiclassical Bekenstein-Hawking entropy of certain black hole solutions in string theory\cite{Dijkgraaf,levsar}.

This wide variety of
physical applications
justifies an attempt to present a self contained
entanglement based reformulation of the results on the classification of three-forms in $6,7,8,9$ dimensions.
Apart from shedding new light on special fermionic systems and presenting some of their invariants serving as measures of entanglement in a unified manner, this approach
also facilitates an embedding of special entangled systems of distinguishable constituents like three-qubits and three-qutrits.
In this philosophy systems with distinguishable constituents are just special cases of systems with indistinguishable ones.

For clarity we would like to note that the methods presented here are not directly applicable when one considers entanglement between modes\cite{Zanardi,Banuls,Heaney} of indistinguishable systems. Mode entanglement is particulary usefull when one wants to classify entanglement between different momenta or different regions of space. However, entanglement in this notion involves the splitting of fermionic mode operators $f_i^\dagger$ into subsets which is not invariant under local unitary transformations of the form $f_i^\dagger \mapsto U_i^{\;j}f_j^\dagger$ (e.g. the Fourier transformation on a lattice) which is a key ingredient in conventional entanglement classification between particles.

This paper is organized as follows. In Section \ref{sec:multilin}. we give a brief introduction to the language of multilinear algebra for the reader unfamiliar with it. This language turns out to be a particulary useful tool for generating SLOCC covariants and invariants. In Section \ref{sec:covariants}. we introduce a family of linear maps or \textit{covariants} derived from the amplitudes of a fermionic state. All the invariants considered in this paper are derived from this construction. In Section \ref{sec:threefermions}. we present the SLOCC classification for three fermion systems in dimensions 6, 7, 8 and 9. 
For the six and seven dimensional cases we present the structure of the SLOCC classes in a new manner based on the structure of the Fano plane. 
In addition to a discussion of the SLOCC classes we present all the algebraically independent continuous invariants of these systems. Most of these invariants are known and used in different fields of physics and mathematics although except for the case of 6 dimensions, they have not made their debut in quantum information theory yet. 
We also discuss the embedding of three qubits into the system of three fermions with six single particle states and show how the measures of entanglement are related. There is a similiar possibility of embedding three qutrits into the system of three fermions with nine single particle states. We consider this case in Section \ref{sec:3qutrits}. and relate the invariants of three qutrits to the ones of the corresponding fermionic system. In Section \ref{sec:pinning}. we outline some of the connections of these results with the entanglement polytopes of Klyachko in particular with the pinning of fermionic occupation numbers which is a concept of huge interest recently\cite{Christandl2}.
Our conclusions are left to Section \ref{sec:conclusions}.
For the convenience of the reader we included two Appendices with some proofs and calculational details.

\section{Multilinear algebra}
\label{sec:multilin}

In this section we give a brief summary of the language of
multilinear algebra which is a useful tool for attacking the
entanglement classification problem of multifermion systems. The
reader familiar with these concepts may skip to the next section.

Let $V\cong\mathbb{C}^N$ be an $N$ dimensional complex vector space. Denote the Cartesian product of $V$ with itself by $V\times V$. There are two canonical ways of defining a vector space from $V\times V$. The first is the direct product the second is the direct sum. The direct product of vectors is defined by the relations $(v+u)\otimes w=v\otimes w+u\otimes w$, $v\otimes(u+w)=v\otimes u + v\otimes w$, $(cv)\otimes w=v\otimes (cw)=c(v\otimes w)$ where $u,v,w\in V$, $c\in \mathbb{C}$. The vector space spanned by elements of the form $v\otimes w$ is denoted by $V\otimes V$ or $V^{\otimes 2}$. If $\lbrace e_i \rbrace_{i=1}^N$ is a basis in $V$ then $\lbrace e_i \otimes e_j\rbrace_{i,j=1}^N$ is a basis of $V\otimes V$. Obviously $V\otimes V$ has dimension $N^2$. Similiary one can
define the $k$th tensor power of $V$ denoted by $V^{\otimes k}$ spanned
by elements of the form $v_1\otimes v_2 \otimes ...\otimes v_k$.
This has dimension $N^k$.
The tensor product is now a map $\otimes:\; V^{\otimes k}\times V^{\otimes m}\rightarrow V^{\otimes(k+m)}$.

The wedge product of $k\leq N$ vectors is defined as \beq v_1
\wedge ... \wedge v_k = \frac{1}{k!}\sum_{\pi\in S_k}
\sigma(\pi)v_{\pi(1)}\otimes ... \otimes v_{\pi(k)}, \eeq where
$S_k$ is the symmetric (permutation) group and $\sigma$ is its
alternating representation, namely $\sigma(\pi)=1$ for even,
$\sigma(\pi)=-1$ for odd permutations. The vector space spanned by
elements of the form $v_1 \wedge ... \wedge v_n$ is denoted by
$\wedge^k V$ and has dimension $\binom{N}{k}$. Its elements are denoted with
$\alpha,\beta,\gamma,\dots $ and we will call them $k$-vectors.

The direct sum is defined from $V\times V$
with the relations $(v+u)\oplus(w+z)=v\oplus w + u\oplus z$, $(cv)\oplus (cw)=c(v\oplus w)$. The vector space obtained in this way is denoted by $V\oplus V$. If $\lbrace e_i \rbrace_{i=1}^N$ is a basis of $V$ then $\lbrace e_i\oplus 0, 0\oplus e_i\rbrace_{i=1}^N$ is a basis in $V\oplus V$ (here 0 denotes the zero vector in $V$). Thus the dimension of $V\oplus V$ is simply $2N$.

Define now the vector space
\beq
\wedge(V)=\mathbb{C}\oplus V\oplus \wedge^2 V \oplus ... \oplus \wedge^N V.
\eeq
Now $\wedge(V)$ can be elevated into an algebra via extending linearly the exterior product
\beq
\begin{aligned}
\wedge : \wedge(V)\times \wedge(V) &&\rightarrow \wedge (V), \\
\alpha, \beta &&\mapsto \alpha \wedge \beta.
\end{aligned}
\eeq
Endowed with this product $\wedge(V)$ is called an exterior algebra or Grassman algebra. The exterior product is a graded anticommutative product, meaning that for $\alpha \in \wedge^k V$ and $\beta \in \wedge^m V$ we have
\beq
\alpha \wedge \beta = (-1)^{km}\beta\wedge \alpha.
\eeq
Fixing a basis $\lbrace e_i\rbrace_{i=1}^N$ in $V$ allows one to write $\alpha\in \wedge^kV$ in the form
\beq
\alpha = \frac{1}{k!}\alpha^{i_1...i_k}e_{i_1}\wedge...\wedge e_{i_k},
\eeq
where $\alpha^{i_1...i_k}$ is totally antisymmetric in all of its indices and summation for the indices is understood.

Let $V^*$ be the dual space of $V$ comprising the linear
functionals acting on $V$. If $\lbrace e_j\rbrace_{j=1}^N$ refers
to a basis of $V$ and $\lbrace e^i\rbrace_{i=1}^N$ a basis of
$V^{\ast}$ then we have $\langle e^i,e_j\rangle
=\delta^i_{\;\;j}$. One can also define the exterior algebra of
$V^*$ denoted by $\wedge(V^*)$. Its elements $P,Q,R,\dots$ will be
called $k$-forms. An element $P$ of $\wedge^k V^*\cong (\wedge^k
V)^*$ is a multilinear functional $P: V\times ...\times
V\rightarrow \mathbb{C}$ on $V$ satisfying
$P(v_1,...,v_k)=\sigma(\pi)P(v_{\pi(1)},...,v_{\pi(k)})$ for all
$\pi\in S_k$. A general element $P \in \wedge^k V^*$ can be
written as \beq P =
\frac{1}{k!}P_{i_1...i_k}e^{i_1}\wedge...\wedge e^{i_k}.
\label{kforma}\eeq The pairing $\langle\cdot,\cdot\rangle$ between
one-forms and vectors gives rise to a natural pairing between
$k$-forms and $k$-vectors. In terms of basis vectors it reads \beq
\langle e^{i_1}\wedge\cdots e^{i_k},e_{j_1}\wedge\cdots
e_{j_N}\rangle ={\rm
Det}\begin{pmatrix}\delta^{i_1}_{\;\;j_1}&\cdot&\cdot&\delta^{i_1}_{\;\;j_N}\\
\cdot&\cdot&\cdot&\cdot&\\
\cdot&\cdot&\cdot&\cdot&\\
\delta^{i_N}_{\;\;j_1}&\cdot&\cdot&\delta^{i_N}_{\;\;j_N}\end{pmatrix}.
 \label{pairingforforms}\eeq\noindent

There is a useful structure connecting the exterior algebra and
its dual, called the interior product or contraction. For a vector
$v\in V$ the interior product ${\iota}_v$ is a
$\wedge^kV^{\ast}\to \wedge^{k-1}V^{\ast}$ linear mapping given by
the defining formula \beq {\iota}_ve^{i_1}\wedge\cdots\wedge
e^{i_k}=\sum_{n=1}^k(-1)^{k-1}\langle e^{i_n},v\rangle
e^{i_1}\wedge\cdots\wedge \check{e}^{i_n}\wedge\cdots \wedge
e^{i_k}.\label{interiordef}
\eeq \noindent 
Where the notation $\check{e}^{i_n}$ means that $e^{i_n}$ has to be omitted from the product. 
For a $k$-form $P$
having the form (\ref{kforma}) we have the
explicit expression for the contraction:
\beq
{\iota}_vP=\frac{1}{(k-1)!}v^{i_1}P_{i_1i_2\dots
i_k}e^{i_2}\wedge\cdots\wedge e^{i_k}.\label{intexplicit} 
\eeq
\noindent The definition of the contraction is a natural notion
justified by the important identity \beq \langle
{\iota}_{e_a}e^{i_1}\wedge e^{i_2}\cdots\wedge
e^{i_k},e_{j_2}\wedge\cdots\wedge e_{j_k}\rangle =\langle
e^{i_1}\wedge e^{i_2}\wedge\cdots\wedge e^{i_k},e_a\wedge
e_{j_2}\wedge\cdots\wedge e_{j_k}\rangle.
\label{fontosid}\eeq\noindent This definition can be extended by
linearity to a one featuring a contraction by an arbitrary
$m$-vector.
 \beq
\begin{aligned}
\label{eq:interior}
\iota: \wedge^m V \times \wedge^k V^* &\rightarrow \wedge^{k-m}V^* \\
  \alpha &=\frac{1}{m!}\beta^{i_1...i_m}e_{i_1}\wedge ...\wedge e_{i_m}, \\ P &=\frac{1}{k!}P_{i_1...i_k}e^{i_1}\wedge ...\wedge e^{i_k} \\
 \alpha,P &\mapsto \iota_\alpha P= \frac{1}{(k-m)!}\alpha^{i_1...i_m}P_{i_1...i_m i_{m+1}...i_k}e^{i_{m+1}}\wedge ... \wedge e^{i_k}\in \wedge^{k-m}V^*,
\end{aligned}
\eeq
where $k\geq m$. We have the useful properties:
\beq
\label{eq:interiorprop}
\begin{aligned}
\iota_\alpha \circ \iota_\beta = (-1)^{km}\iota_\beta \circ \iota_\alpha, && \alpha \in \wedge^k V,\; \beta\in \wedge^m V,\\
\iota_\alpha (P \wedge Q ) = \iota_\alpha (P)\wedge Q + (-1)^{kp}P\wedge \iota_\alpha (Q), && \alpha\in \wedge^k V, \; P \in \wedge^p V^*, \; Q \in \wedge^q V^*,\\
&& k\leq p,q.
\end{aligned}
\eeq

There is an important isomorphism relating $m$-forms and $N-m$
vectors. It reads
 \beq  \wedge^m V^* \cong
\wedge^{N-m}V\otimes \wedge^N V^*.\label{isofontos} \eeq\noindent
This isomorphism is based on the definition of the $\star$
operation defined as follows.  \beq Q \wedge R = \langle Q,\star
R\rangle\qquad Q\in \wedge^{N-m} V^*,\qquad R \in \wedge^m V^*,
\qquad \star R\in \wedge^{N-m}\otimes\wedge^NV^{\ast}.
\label{definingstar}\eeq\noindent Using the
(\ref{pairingforforms}) identity one can show that \beq\star
R=\frac{1}{(N-m)!}(\star R)^{i_1\cdots
i_{N-m}}e_{i_1}\wedge\cdots\wedge
e_{i_{N-m}}\otimes\mathbb{E}\label{csillagvektor}\eeq \noindent
where \beq (\star R)^{i_1\cdots
i_{N-m}}=\frac{1}{m!}\varepsilon^{i_1\cdots i_{N-m}j_1\cdots
j_m}R_{j_1\cdots j_m},\qquad \mathbb{E}=e^1\wedge\cdots\wedge e^N.
\label{igynezki} \eeq\noindent It should be emphasized that
$\star$ is not the Hodge star, until this point we did not
equip $V$ with any metric.

Let $g=g^j_{\;\;i}e^i\otimes e_j\in GL(V)$ be an invertible linear
map from $V$ to itself acting on a $v\in V$ as $g
v=g^j_{\;\;i}v^k\langle e^i,e_k\rangle\otimes e_j=g^j_{\;\;k}v^k
e_j$. For this action on the basis vectors we write  
\beq g
e_i=e_jg^j_{\;\;i},\quad g\in GL(V). \label{vektorhatas} 
\eeq
\noindent Given this action on $V$ an action $g^{\ast}$ on
$V^{\ast}$ is induced via the formula 
\beq \langle g^{\ast} e^i, g
e_j\rangle= \langle e^i,e_j\rangle
=\delta^i_{\;\;j}.\label{dualhatas} 
\eeq\noindent 
Explicitly we
have 
\beq 
g^{\ast} e^i=e^j{g^{\prime}}_j^{\;\;i},\qquad
{g^{\prime}}_k^{\;\;i}g^k_{\;\;j}={\delta^i}_j \label{expldualhat}
\eeq \noindent 
i.e. the matrix of $g^{\prime}$ is just the inverse
transpose of the matrix of $g$ \beq g^{\prime}=(g^t)^{-1}.
\label{transposeinverse}\eeq \noindent
 Now this dual action induces an action $\varrho(g)$
 on $\wedge^k V^*$.
 However, by an abuse of notation we use again $g^{\ast}$ for this action
 \beq g^*:\wedge^k V^{\ast} \rightarrow
\wedge^k V^{\ast},\qquad P \mapsto g^{\ast} P.\label{sloccaction}
\eeq \noindent
 For
the components this reads as \beq  P_{i_1...i_k} \mapsto (g^{\ast}
P)_{i_1...i_k}={g^{\prime}}_{i_1}^{\;\;{j_1}}
{g^{\prime}}_{i_2}^{\;\;{j_2}}\cdots
{g^{\prime}}_{i_k}^{\;\;{j_k}}P_{j_1...j_k}.\label{kformakomphatas}
\eeq \noindent Similarly the action on the components of a
$k$-vector $\alpha$ reads as \beq \alpha^{i_1...i_k} \mapsto
(g\alpha )^{i_1...i_k}={g}^{i_1}_{\;\;{j_1}}
{g}^{i_2}_{\;\;{j_2}}\cdots
{g}^{i_k}_{\;\;{j_k}}\alpha^{j_1...j_k}.\label{kvektorkomphatas}
\eeq \noindent By virtue of Eq.(\ref{expldualhat}) in the special
case of the top form $\mathbb{E}$ we have the transformation
formula \beq g^{\ast}\mathbb{E}=({\rm
Det}g)^{-1}\mathbb{E}.\label{toptrans}\eeq \noindent

\section{SLOCC invariants for fermionic systems}
\label{sec:covariants}

Now let us identify $V=\mathbb{C}^N$ with the finite dimensional
single particle Hilbert space. The full Hilbert space of a system
with an indefinite number of fermions is called the Fock space.
Let us denote the vacuum state of the Fock space as $\vert
0\rangle$. Let us moreover define the fermionic operators $f_i$,
$f_j^{\dagger}$ as the ones satisfying the canonical anticommutation
relations 
\beq \{f_i,f^{\dagger}_j\}=\delta_{ij}, \quad
\{f_i,f_j\}=\{f^{\dagger}_i,f^{\dagger}_j\}=0, \quad i,j=1,\dots
N. \label{antika}
\eeq \noindent
 Then the Fock space is spanned by vectors of the form
$f^{\dagger}_{i_1}f^{\dagger}_{i_2}\cdots f^{\dagger}_{i_k}\vert
0\rangle$ with $k=0,1,2,\dots N$.

Now this space can alternatively be represented\cite{Takhtajan} as
the exterior algebra $\wedge(V)$ or $\wedge(V^{\ast})$. For later
convenience we chose $\wedge(V^{\ast})$. In this picture the
operators $f_i$ and $f_i^{\dagger}$ acting on the Fock space are
mapped to the ones $e^i\wedge $ and $\iota_{e_i}$ acting on
$\wedge(V^{\ast})$. If we use $P\in \wedge^kV^{\ast}$ of
(\ref{kforma}) as the representative of the unnormalized
$k$-fermion state  \beq \vert P\rangle =P_{i_1i_2\cdots
i_{k}}f^{\dagger}_{i_1}f^{\dagger}_{i_2}\cdots
f^{\dagger}_{i_k}\vert 0\rangle \label{kfermionstate} \eeq
\noindent then the action of the fermionic operators on the usual
Fock space can be represented as the
 \beq
 \label{eq:keltowedgemegf}
\begin{aligned}
f_i^\dagger |P\rangle &&\mapsto e^i\wedge P, \\
f_i |P\rangle &&\mapsto \iota_{e_i} P
\end{aligned}
\eeq action on $k$-forms. This map clearly gives a representation
of the (\ref{antika}) anticommutation relations. Indeed from \eqref{eq:interiorprop} one sees that
\beq
\iota_{e_i}(e^j\wedge P)=\delta^j_{\;i}P-e^j\wedge (\iota_{e_i} P),
\eeq
hence $\lbrace \iota_{e_i}, e^j\wedge\rbrace=\delta^j_{\;i}$.

Let then $V$ be an $N$ dimensional complex vector space
representing the one particle states of a fermionic system and the
unnormalized $k$ fermion states be represented as in
Eq.(\ref{kforma}). The $P_{i_1...i_k}$ in this formula are the
$\binom{N}{k}$ complex amplitudes characterizing the $k$ fermion
state. Here we are dealing with a system of indistinguishable
constituents hence SLOCC transformations are acting via the same
$GL(V)=GL(N,\mathbb{C})$ map on each slot as defined in
\eqref{kformakomphatas}.

Two fermionic states $P$ and $P^{\prime}$ are called SLOCC
equivalent if there exists an element of $g\in GL(V)$ such that
$P^{\prime}=g^{\ast}P$. The abbreviation SLOCC refers to
stochastic local operations and classical
communication\cite{Bennett,Dur} the type of physical manipulations
represented mathematically by invertible linear transformations
$g\in GL(V)$. Sometimes the unimodular subgroup
$SL(V)=SL(N,\mathbb{C})$ is also used to define new equivalence
classes. The subgroup \beq {\rm Stab}(P)=\{g\in GL(V)\vert
g^{\ast}P=P\} \label{stab} \eeq \noindent is called the stabilizer
subgroup of the multifermion state. Under the SLOCC equivalence
relation one can form the corresponding equivalence classes. We
will refer to these classes as the SLOCC {\it entanglement
classes}.

In order to distinguish between different types (classes) of
entanglement one can introduce entanglement measures. An
entanglement measure is a real valued function $f$ of the
amplitudes $P_{i_1\dots i_k}$ satisfying a number of physically
useful properties\cite{Plenio}. Here we will be content merely
with one of such properties, namely that our measures should be
coming from relative invariants under the SLOCC group, (invariants
under the unimodular SLOCC group). A rational function $I:
\wedge^kV^{\ast}\to \mathbb{C}$ is a SLOCC relative invariant if
there exists a rational character $\chi:GL(V)\to GL(1,\mathbb{C})$
, i.e. a one-dimensional rational representation such that \beq
I(g^{\ast}P)=\chi(g)I(P). \label{relinv} \eeq \noindent If
$\chi\equiv 1$ then $I$ is called  an invariant. The entanglement
measures $f:\wedge^{\ast}V\to\mathbb{R}$ studied here are arising
as magnitudes of relative invariants with respect to the SLOCC
group (invariants under the unimodular SLOCC group).

There is a number of covariants that can be defined to form such invariants.
For a study of covariants and invariants useful in the
fermionic context see the book of Gurevich\cite{Gurevich}. Here we will content with some of his constructions
suitably modified and adapted to our purposes.
For a multifermionic state $P$
one can review a collection of SLOCC invariants as follows.

\textbf{Degree 1 invariants:} These are the ranks of the linear maps that can be constructed from $P$ and are linear in the amplitudes. Let $\iota$ denote the interior product of equation \eqref{eq:interior}.
Define the set of linear maps
\beq
\begin{aligned}
P^{(l)}: \wedge^l V& \rightarrow \wedge^{k-l}V^*\\
 \alpha& \mapsto {\iota}_{\alpha} P.
\end{aligned}
\label{kovi1} \eeq Now $P^{(l)}$ is a linear map from a vector
space of dimension $\binom{N}{l}$ to a vector space of
$\binom{N}{k-l}$ thus it has a SLOCC invariant rank at most
min$(\binom{N}{l},\binom{N}{k-l})$. However, not all of these are
independent. Obviously $P^{(k-l)}$ is the transpose of $P^{(l)}$
thus their rank is equal.

\textbf{Degree 2 invariants:} These are ranks of linear maps wich are quadratic in the amplitudes of $P$. Let
\beq
\label{eq:kappa}
\begin{aligned}
{\tilde \kappa}_P^{(l)}: \wedge^{l}V &\rightarrow \wedge^{2k-l}V^* \\
\alpha &\mapsto {\iota}_{\alpha}P\wedge P.
\end{aligned}
\eeq Now by virtue of the (\ref{isofontos}) isomorphism one can
define a new quantity
 \beq
 \kappa^{(l)}_P\equiv\star \circ {\tilde \kappa}^{(l)}_P\label{kappaujabb}\eeq\noindent
 which is a \textit{linear} map from
$\wedge^{l}V$ to $\wedge^{N-2k+l}V \otimes \wedge^N V^*$. The
appearance of the one dimensional space $\wedge^N V^*$ means that
according to Eq.(\ref{toptrans}) this object picks up a
determinant factor under a SLOCC transformation. Obviously this
construction only makes sense if $0\leq l\leq k$ satisfies \beq
0\leq 2k-l \leq N. \eeq Let us give the explicit form of
$\kappa^{(l)}_P(\alpha)$ \beq
\kappa^{(l)}_P(\alpha)=\frac{1}{(N-2k+l)!}{(K_P^{(l)})^{a_1\cdots
a_{N-2k+l}}}_{b_1\cdots b_l}\alpha^{b_1\cdots
b_l}e_{a_1}\wedge\cdots\wedge e_{a_{N-2k+l}}\otimes\mathbb{E}
\label{jobonyolult} \eeq \noindent where

\beq {(K_P^{(l)})^{a_1\cdots a_{N-2k+l}}}_{b_1\cdots
b_l}=\frac{1}{(k-l)!k!}\varepsilon^{a_1\cdots a_{N-2k+l}i_1\cdots
i_{k-l}i_{k-l+1}\cdots i_{2k-l}}P_{b_1\cdots b_li_1\cdots
i_{k-l}}P_{i_{k-l+1}\cdots i_{2k-l}}. \label{Kexplicit} \eeq
\noindent Clearly the $\binom{N}{2k-l}\times\binom{N}{l}$ matrices
$K^{(l)}_P$ have a SLOCC invariant rank. The index structure of
$K_P^{(l)}$ shows that under SLOCC transformations the upper
indices are transformed via the use of $N-2k+l$ matrices
$g^i_{\;\;j}$ and the lower indices via the use of $l$ matrices
${g^{\prime}}_i^{\;\;j}$, moreover due to the presence of the
Levi-Civit\'a symbol (compare also with the transformation rule of
Eq. (\ref{toptrans})) an extra factor of ${\rm
Det}g^{\prime}=({\rm Det}g)^{-1}$ appears.

\begin{prop}
$\tilde{\kappa}^{(k-1)}_P=0$ if and only if $P$ is separable.
\begin{proof}
Let $\alpha \in \wedge^{k-1}V$. By definition \beq
\tilde{\kappa}^{(k-1)}_P(\alpha)=\alpha^{i_1...i_{k-1}}P_{i_1...i_{k-1}i_k}
P_{j_1...j_k}e^{i_k}\wedge e^{j_1}\wedge  ...\wedge e^{j_k}. \eeq
Since $\alpha$ is arbitrary then our condition reads as \beq
P_{i_1...i_{k-1}[i_k} P_{j_1...j_k]}=0, \label{Penrose} \eeq
\noindent where the brackets denote antisymmetrization. It can be
shown (see e.g. Proposition 3.5.30 of the book of Penrose and
Rindler\cite{Penroserindler}) that Eq.(\ref{Penrose}) is a
sufficient and necessary condition for $P_{i_1\cdots i_k}$ to be
separable i.e. of the form $P_{i_1\cdots
i_k}=a_{[{i_1}}b_{i_2}\cdots z_{{i_k}]}$. 
\end{proof}
\end{prop}
These amplitudes can be
expressed in terms of a single Slater determinant hence they
represent separable multifermion states. Note that for these
sufficient and necessary conditions of separability an equivalent
form is provided by the set of Pl\"ucker relations usually
expressed\cite{Kasman} in the 
\beq 
\Pi_{\cal{ A},\cal{B}}
=\sum_{n=1}^{k+1}(-1)^{n-1}P_{i_1i_2\dots
i_{k-1}j_n}P_{j_1j_2\dots j_{k+1}\hat{j}_n}=0, \label{general}
\eeq \noindent
form. Here ${\cal A}=\{i_1,i_2,\dots ,i_{k-1}\}$
and ${\cal B}=\{j_1,j_2,\dots ,j_{k+1}\}$ are $k-1$ and $k+1$
element subsets of the set $\{1,2,\dots, N\}$, and where the
number $\hat{j}_n$ has to be omitted .

\textbf{Degree $n+1$ invariants:} Define \beq \label{eq:kappa}
\begin{aligned}
\kappa_P^{(l_1 l_2 ... l_n)}: \otimes_{j=1}^n (\wedge^{l_j}V) &\rightarrow \wedge^{k+nk-\sum_{j=1}^n l_j}V^* \cong \wedge^{N-k(n+1)+\sum_{j=1}^n l_j}V \otimes \wedge^N V^*, \\
\alpha_1,...,\alpha_n &\mapsto  \star\left(\iota_{\alpha_{1}}P\wedge ... \wedge \iota_{\alpha_{n}}P \wedge P\right), \\
\alpha_i &\in \wedge^{l_i}V.
\end{aligned}
\eeq Just like the ones of Eq. (\ref{jobonyolult}) these
quantities are based on $\prod_{j=1}^n \binom{N}{l_j}$ times
$\binom{N}{(n+1)k-\sum_{j=1}^n l_j}$ matrices $K_P^{(l_1\cdots
l_n)}$ with a SLOCC invariant rank. For the definition to make
sense, we have the constaint for the $0\leq l_j\leq k$: \beq
\label{eq:constr} 0\leq (n+1)k-\sum_{j=1}^n l_j\leq N. \eeq These
covariants with degree over 2 can have extra symmetry properties
if there exists $l_i=l_j$ for some $i\neq j$. Consider for example
$\kappa^{(ll)}_P$. Then we have \beq
\kappa^{(ll)}_P(\alpha_1,\alpha_2)=(-1)^{k-l}\kappa^{(ll)}_P(\alpha_2,\alpha_1),
\;\;\alpha_1,\alpha_2\in \wedge^l V. \eeq

As an example needed later on let us consider the special case of
${\kappa}_P^{(l_1\cdots l_n)}$ with $l_1=\cdots =l_n=1$ for three
fermion systems with $N$ single particle states. In this case
$k=3$, $\alpha=\alpha^be_b\in V$ and we define $m$ via $3+2n+m=N$.
For simplicity in this case we will refer to
${\kappa}_P^{(l_1\cdots l_n)}$ as ${\kappa}_P^{[m,n]}$.
 Then ${\kappa}_P^{[m,n]}
\wedge^n V \rightarrow \wedge^mV\otimes \wedge^NV^{\ast}$ is having the form \beq
{\kappa}_P^{[m,n]}(\alpha_1,\cdots,\alpha_n)=\frac{1}{m!}{\left(K_P^{[m,n]}\right)^{a_1\cdots
a_m}}_{b_1\cdots b_n}{\alpha_1}^{b_1}\cdots
{\alpha_n}^{b_n}e_{a_1}\wedge\cdots\wedge
e_{a_m}\otimes\mathbb{E}\label{megrondabb} \eeq \noindent where
\beq  {\left(K_P^{[m,n]}\right)^{a_1\cdots a_m}}_{b_1\cdots b_n}
=\frac{1}{2^n3!}\varepsilon^{a_1\cdots a_mi_1\cdots
i_{2n+3}}P_{b_1i_1i_2}\cdots
P_{b_ni_{2n-1}i_{2n}}P_{i_{2n+1}i_{2n+2}i_{2n+3}}.\label{nagykmatrixalakja}
\eeq \noindent \ Notice that ${\left(K_P^{[m,n]}\right)^{a_1\cdots
a_m}}_{b_1\cdots b_n}$ is totally antisymmetric in its upper, and
symmetric in its lower indices.

The ranks of the linear maps defined above
are SLOCC invariants because a SLOCC
transformation on them simply means an
invertible change of basis in the domain and
the range and a multiplication by some power of the SLOCC determinant.
However, these ranks are not continuous invariants in the amplitudes $P_{i_1i_2i_3}$. We can also use the above defined linear maps to define continuous relative SLOCC invariants.
The idea is to utilize the trace and determinant defined on
linear automorphisms of vector spaces.
In order to do this we need to construct square matrices.
This can be done by composing maps with each other to have the
same dimensional range and domain.
As we will see the simplest case arises when the above defined maps are square matrices themselves.

It is also worth noting that a system of $k$ qudits with Hilbert
space $\mathcal{H}=\mathbb{C}^d\otimes ... \otimes\mathbb{C}^d$
can be embedded in this special fermionic system\cite{levvran2} in
the following way 
\beq
\begin{aligned}
|\psi\rangle &=\sum_{\mu_1,...,\mu_k=1}^d \psi_{\mu_1...\mu_k}|\mu_1\rangle\otimes...\otimes |\mu_k\rangle \in \mathcal{H}\\
P_\psi &= \sum_{\mu_1,...,\mu_k=1}^d \psi_{\mu_1...\mu_k} e^{\mu_1}\wedge e^{d+\mu_2} \wedge ...\wedge e^{(k-1)d+\mu_k}.
\end{aligned}
\label{eq:quditembed} 
\eeq
Obviously a SLOCC transformation on
$\mathcal{H}$ of the form $g_1\otimes...\otimes g_k \in
GL(d,\mathbb{C})^{\otimes k}$ acting on $\psi$ like \beq
\psi_{\mu_1...\mu_k} \mapsto
(g_1)_{\mu_1}^{\;\nu_1}...(g_k)_{\mu_k}^{\;\nu_k}\psi_{\mu_1...\nu_k}\label{szokasoshatas}
\eeq can be embedded in the SLOCC group $GL(V)$ of our fermionic
system via \beq g= \left(
\begin{array}{ccc}
g_1 & & \\
& \ddots& \\
& & g_k
\end{array}
\right) \in GL(dk,\mathbb{C})=GL(V). \label{szokasosbeagy} \eeq As
a consequence embedded states on different $GL(V)$ orbits must be in
different $GL(d,\mathbb{C})^{\otimes k}$ orbits as well. However, the
converse is not generally true, entanglement classes of the fermionic system
may split into different classes when just the embedded system is
considered. 
However, when we consider the  generalized SLOCC group i.e. the SLOCC group combined with permutations some important exceptions arise. In the case of three qubits the embedding into three fermions with six single particle states is bijective between the SLOCC classes of the two systems. 
As was shown in the case of four qubits embedded into the system of four fermions with eight single particle states two inequivalent four qubit states remain inequivalent under the fermionic SLOCC group\cite{DjokPRA}. As pointed out in Section \ref{sec:3qutrits}. splitting does not occur between \textit{families} of entanglement classes for the embedding of three qutrits into the system of three fermions with nine single particle states. Most likely this is true for the entanglement classes too.	
In the cases when splitting of fermionic entanglement classes does occur one can still use the ranks of the maps
$\kappa^{(l_1...l_n)}_{P_\psi}$ in order to obtain  a coarse-graining
of the entanglement classes of $\mathcal{H}$.

Finally note that one can see from the isomorphism \eqref{eq:keltowedgemegf} that the anticommutation relations \eqref{antika} are invariant under invertible SLOCC transformations. Based on this property one can extend the group $GL(V)$ acting on fermionic states to a bigger one which also enables the implementation of Bogoliubov transformations. This way one can obtain a classification of states on the whole fermionic Fock space not just on the fixed particle number subspaces. For details on this subject see our recent work\cite{levsar2}.

\section{Entanglement of three fermions}
\label{sec:threefermions}

\subsection{Six single particle states}
\label{sec:threefsixs}

The entanglement classification of three fermions with six single
particle states is already well known and has a broad connection
with several mathematical and physical structures\cite{Borland,
Christandl,Christandl2,Djok2} in the literature. It was first
recognized as a QIT problem in Ref.\cite{levvran1}, where also the
connection to Freudenthal triple systems has been revealed. Later
it has been realized that the corresponding mathematical problem
has already been solved long ago\cite{Reichel} and that the
generic SLOCC orbit is precisely the one which shows up in the
theory of prehomogenous vector spaces\cite{Satokimura,Kimura}.
Moreover, within such three fermionic systems three-qubit systems
can be embedded\cite{levvran1,levvran2,Djok2,DjokPRA} in this case this generic
SLOCC class corresponds to the famous GHZ-class\cite{Dur} of
three-qubit entanglement.  Furthermore, recently it has been shown
that the problem is even connected to string theory via the so
called Hitchin functionals\cite{stable,Hitchin,levsar}.

Let $V$ be a the six dimensional complex vector space
$\mathbb{C}^6$. Then an unnormalized three fermion state can be
represented as \beq P=\frac{1}{3!}P_{i_1i_2i_3}e^{i_1}\wedge
e^{i_2} \wedge e^{i_3} \in \wedge^3 V^*. \label{3fw6}\eeq The
$P_{i_1i_2i_3}$ are the 20 complex amplitudes describing the three
fermion state. The SLOCC transformations act with the same
$GL(V^*)=GL(6,\mathbb{C})$ map on each slot as \beq
\label{eq:sloccaction} P_{i_1i_2i_3} \mapsto
{g^{\prime}}_{i_1}^{\;\;j_1} {g^{\prime}}_{i_2}^{\;\;j_2}
{g^{\prime}}_{i_3}^{\;\;j_3}
 P_{j_1j_2j_3}
\eeq \noindent
  just as defined in Eqs. (\ref{kformakomphatas}), (\ref{transposeinverse}).

  In the following we show that the SLOCC
orbits of this system are completely characterized by the ranks of
the degree one $P^{(2)}$ (Eq. (\ref{kovi1})) and the degree two
$\kappa^{(l_1=1)}_P=\kappa^{[1,1]}_P$ (Eq. (\ref{jobonyolult})) covariants. In
order to see this let us consider the latter one. According
to Eq.(\ref{Kexplicit}) its underlying $6\times 6$ matrix has the
explicit form 
\beq
\left({K^{[1,1]}_P}\right)^a_{\;b}=\frac{1}{2!3!}
\varepsilon^{ai_1i_2i_3i_4i_5}P_{bi_1i_2}P_{i_3i_4i_5}
\label{Kmatrixamikell}
\eeq 
where we also used the notation
introduced in Eq.(\ref{nagykmatrixalakja}). By construction
$K^{[1,1]}_P$ transforms under SLOCC transformations as 
\beq
\label{eq:kappaslocc} {(K^{[1,1]}_P)}^a_{\;b} \mapsto {\rm
Det}(g^{\prime})g^a_{\;\;c}{g^{\prime}}_b^{\;\;d}{(K^{[1,1]}_P)}^c_{\;d}
 \qquad g \in GL(V).
 \eeq 
 According to Eq.(\ref{transposeinverse}) the
 matrix $g^{\prime}$ is just the
 inverse transpose of the one $g$ hence this transformation rule is of the form
 $K_P^{[1,1]}\mapsto \left({\rm Det}(g)\right)^{-1}
 gK_P^{[1,1]}g^{-1}$. It follows that any power of the trace of $K_P^{[1,1]}$ is a relative invariant.
 One can immediately check that
 ${\rm Tr}K_P^{[1,1]}=0$, hence the next item in the line to experiment with is
 ${\rm Tr}\left(K_P^{[1,1]}\right)^2$.

 It is well known that this quantity suitably normalized  
 \beq
\mathcal{D}(P)=\frac{1}{6}\text{Tr} \left({K^{[1,1]}_P}\right)^2
\label{Threetanglegen} 
\eeq 
is indeed a relative invariant and its
magnitude defines a good measure of entanglement. That
$\mathcal{D}$ is a relative invariant transforming as 
\beq
\mathcal{D}(P)\mapsto({\rm Det}(g^{\prime}))^{2}\mathcal{D}(P).
\label{relativetrafode} 
\eeq 
\noindent can immediately be seen
from the transformation  property of Eq.(\ref{eq:kappaslocc}) and
the definition of Eq.(\ref{Threetanglegen}).
 In order to see the last
property namely that its magnitude provides a measure of
entanglement let us give this relative invariant another
look\cite{levvran1}. First we reorganize the $20$ independent
complex amplitudes $P_{i_1i_2i_3}$ into two complex numbers
$\eta,\xi$ and two complex $3\times 3$ matrices $X$ and $Y$ as
follows. As a first step we change our labelling convention by
using the symbols $\dot{1},\dot{2},\dot{3}$ instead of $4,5,6$
respectively hence we have \beq (1,2,3,4,5,6)\leftrightarrow
(1,2,3,\dot{1},\dot{2},\dot{3}). \label{atnevezes} \eeq \noindent
Hence for example we can alternatively refer to $P_{456}$ as
$P_{\dot{1}\dot{2}\dot{3}}$ or to $P_{125}$ as $P_{12\dot{2}}$.
Now we define \beq \eta\equiv P_{123},\qquad \xi\equiv
P_{\dot{1}\dot{2}\dot{3}} \label{etaxi} \eeq \beq
X=\begin{pmatrix}X_{11}&X_{12}&X_{13}\\X_{21}&X_{22}&X_{23}\\X_{31}&X_{32}&X_{33}\end{pmatrix}
\equiv\begin{pmatrix}P_{1\dot{2}\dot{3}}&P_{1\dot{3}\dot{1}}&P_{1\dot{1}\dot{2}}\\
P_{2\dot{2}\dot{3}}&P_{2\dot{3}\dot{1}}&P_{2\dot{1}\dot{2}}\\P_{3\dot{2}\dot{3}}&P_{3\dot{3}\dot{1}}&P_{3\dot{1}
\dot{2}}\end{pmatrix}, \label{ymatr} \eeq \beq
Y=\begin{pmatrix}Y_{11}&Y_{12}&Y_{13}\\Y_{21}&Y_{22}&Y_{23}\\Y_{31}&Y_{32}&Y_{33}\end{pmatrix}\equiv
\begin{pmatrix}P_{\dot{1}23}&P_{\dot{1}31}&P_{\dot{1}12}\\P_{\dot{2}23}&P_{\dot{2}31}&P_{\dot{2}12}\\
P_{\dot{3}23}&P_{\dot{3}31}&P_{\dot{3}12}\end{pmatrix}.
\label{Xmatr} \eeq With this notation the quartic polynomial of
Eq.(\ref{Threetanglegen}) is \beq {\cal D}(P)=[\eta\xi -{\rm
 Tr}(XY)]^2-4{\rm Tr}(X^{\sharp}Y^{\sharp})+4\eta{\rm
 Det}(X)+4\xi{\rm Det}(Y), \label{Cayleygen}
 \eeq
 \noindent where $X^{\sharp}$ and $Y^{\sharp}$ correspond to the
 regular adjoint matrices for $X$ and $Y$ hence for example
 $XX^{\sharp}=X^{\sharp}X={\rm Det }(X)I$
 with $I$ the $3\times 3$ identity matrix.

Now according to Eq.(\ref{eq:quditembed}) we can embed a three-qubit
state $\psi$ into our system of three fermions with six single
particle states as the state $P_{\psi}$. 
However, for convenience we choose another form of this embedding\cite{levvran1} which amounts to a permutation $(3245)$ of the basis vectors $e^1,\dots e^6$.
One can show that under this permutation the matrix of embedded SLOCC transformations familiar from Eq.(\ref{szokasosbeagy})
takes a form of a $6\times 6$ matrix consisting of four blocks of $3\times 3$ diagonal matrices.
Via this embedding 
we keep merely $8$ complex amplitudes
from the $20$ ones of $P$ which transform according to the
restricted SLOCC group 
as the amplitudes
of a three-qubit system. Let us label the $8$ amplitudes of $P_{\psi}$ as \beq
(P_{123},P_{12\dot{3}},P_{1\dot{2}3},P_{\dot{1}23},
P_{\dot{1}\dot{2}\dot{3}},P_{\dot{1}\dot{2}3},P_{\dot{1}2\dot{3}},P_{1\dot{2}\dot{3}})
=(\psi_{000},\psi_{001},\psi_{010},\psi_{100},\psi_{111},\psi_{110},\psi_{101},\psi_{011}),
\label{specampl} \eeq \noindent
where unlike in Eq.(\ref{eq:quditembed}) now we switched to the use of the conventional labelling $\mu_1,\mu_2,\mu_3=0,1$ of basis states.
Then $\mathcal{D}(P_{\psi})\equiv
D(\psi)$ takes the following form
\begin{eqnarray}
D(\psi)&=&
[\psi_0\psi_7-\psi_1\psi_6-\psi_2\psi_5-\psi_3\psi_4]^2-
4[(\psi_1\psi_6)(\psi_2\psi_5)+(\psi_2\psi_5)(\psi_3\psi_4)\\\nonumber&+&
(\psi_3\psi_4)(\psi_1\psi_6)]+
4\psi_1\psi_2\psi_4\psi_7+4\psi_0\psi_3\psi_5\psi_6 \label{Cayley}
\end{eqnarray}
where $(\psi_0,\psi_1,\dots,\psi_7)\equiv
(\psi_{000},\psi_{001},\dots,\psi_{111})$. $D(\psi)$ gives rise to
a famous entanglement measure\cite{CKW} called the {\it
three-tangle} $\tau_{123}$ which for normalized states satisfies
\beq 0\leq {\tau}_{123}=4\vert D(\psi)\vert\leq 1.
\label{threetangle} \eeq Hence $\mathcal{D}(P)$ with the
normalization as given by Eq.(\ref{Threetanglegen}) is a natural
generalization of the three-tangle for three fermions with six
single particle states. For normalized fermionic states it can be
shown\cite{levvran1} that an analogous quantity
$\mathcal{T}_{123}$ formed from $\mathcal{D}(P)$ satisfies \beq
0\leq {\cal T}_{123}=4\vert{\cal D}(P)\vert\leq 1
\label{tanglegen} \eeq \noindent just like the three-tangle does
for three-qubits. We note that the expression for $\mathcal{D}$ as
given by Eq.(\ref{Cayleygen}) is just the quartic invariant of the
Freudenthal triple system over the cubic Jordan algebra
$M(3,\mathbb{C})$ of $3\times3$ complex
matrices\cite{levvran1,Kru}.

Let us give yet another form\cite{Freuddual,levsar} of the quartic
invariant $\mathcal{D}(P)$. Define a symplectic form on
$\wedge^3V^{\ast}$  \beq
\{\cdot,\cdot\}:\wedge^3V^{\ast}\times\wedge^3V^{\ast}\to {\mathbb
C},\qquad (P,Q)\mapsto
\frac{1}{3!3!}{\varepsilon}^{ijklmn}P_{ijk}Q_{lmn}, \label{sympl}
\eeq \noindent and a three-form $\tilde{P}$ dual to the original
three-form $P\in \wedge^3V^{\ast}$ as 
\beq
\tilde{P}=\frac{1}{3!}\tilde{P}_{abc}e^a\wedge e^b\wedge e^c,\quad
\tilde{P}_{abc}=\frac{1}{2!3!}{\varepsilon}^{di_2i_3i_4i_5i_6}P_{bcd}P_{ai_2i_3}P_{i_4i_5i_6}
=P_{bcd}{(K_P^{[1,1]})^d}_a. \label{dual} 
\eeq 
\noindent Then the
new form of the quartic invariant is \beq {\cal
D}(P)=\frac{1}{2}\{\tilde{P},P\}. \label{quarticinvper2} \eeq
\noindent In the theory of Freudenthal triple systems the quantity
$\tilde{P}$ which is cubic in the original amplitudes of $P$ is
usually defined via the so called trilinear form\cite{Kru}. With
the help of $\tilde{P}$ for a state with ${\cal D}\neq 0$ one can
define a {\it dual fermionic state} as \beq \hat{P}\equiv
-i\frac{\tilde{P}}{\sqrt {\cal D}}. \label{freudual} \eeq
\noindent With our convention of defining a factor of $-i$ the
expression of $\hat{P}$ boils down to the expression of the so
called {\it Freudenthal dual} of $P$ defined only for {\it real}
states in the paper\cite{Freuddual} of Borsten et.al. One can
check that the dual state satisfies the identities \beq
\mathcal{D}(\hat{P})=\mathcal{D}(P),\qquad \hat{\hat{P}}=-P.
\label{dualisidentitasok} \eeq \noindent Notice also that
according to Eqs.(\ref{relativetrafode}) and (\ref{freudual})
(unlike the quantity $\tilde{P}$) the one $\hat{P}$ does not pick
up a determinant factor under SLOCC transformations.

The classification problem for three-forms in $V={\mathbb C}^6$
under the group action $GL(V)$ has been solved long ago by
Reichel\cite{Reichel}. In the context of fermionic entanglement it
has recently been rediscovered by physicists \cite{levvran1}.
According to this result the $GL(V)$ orbits of three-forms
correspond to the SLOCC orbits of three-fermions with six single
particle states. We have five  SLOCC classes. Using the notation
\beq e^{ijk}\equiv e^i\wedge e^j\wedge e^k\label{jelolestablazat}
\eeq \noindent the representatives of these classes taken together
with the ranks of the basic covariants can be seen in TABLE \ref{tab:1}.

\begin{table}[h!]
\centering
\begin{tabular}{|c|c|c|c|c|}
\hline
Type & Canonical form of $P$ & Rank $P^{(2)}$ & Rank $\kappa^{(1)}_P$ & Rank $\kappa^{(2)}_P$\\
 \hline \hline
Null & 0 & 0 & 0 & 0\\
Sep & $e^{123}$ & 3 & 0 & 0\\
Bisep & $e^{123}+e^{156}$ & 5 & 1 & 4\\
W & $e^{126}+e^{423}+e^{153}$ & 6 & 3 & 6\\
GHZ & $e^{123}+e^{456}$ & 6 & 6 & 6\\  \hline
\end{tabular}
\caption{Entanglement classes of three fermions with six single
particle states, and the ranks of the simplest covariants.}
\label{tab:1}
\end{table}

 The
four nontrivial classes are labelled by the states familiar from
the classification of three-qubits\cite{Dur}. Namely, we have the
totally separable, biseparable, W and GHZ
(Greenberger-Horne-Zeilinger) classes. Using the language of
embedded systems the notation of Eq.(\ref{atnevezes}) and the
mapping of Eq.(\ref{specampl}) one obtains the normalized
representatives of these classes as $\vert 000\rangle$ for the
separable, $(\vert 000\rangle +\vert 011\rangle)/\sqrt{2}$ for the
biseparable, $(\vert 001\rangle +\vert 010\rangle +\vert
100\rangle)/\sqrt{3}$ for the $W$, and $(\vert 000\rangle +\vert
111\rangle)/\sqrt{2}$ for the $GHZ$-class.

An alternative description of the nontrivial SLOCC classes in
terms of normalized representatives can also be given using the
invariant $\mathcal{D}$ and the covariant $\tilde{P}$ as follows.
\begin{equation}
P_{GHZ}=\frac{1}{2}(e^{123}+e^{156}+e^{264}+e^{345}),\quad
\mathcal {D}(P)\neq 0 \label{1v}
\end{equation}
\noindent
\begin{equation}
P_{W}=\frac{1}{\sqrt{3}}(e^{123}+e^{156}+e^{264}),\qquad
\mathcal{D}(P)=0,\quad \tilde P\neq 0 \label{2v}
\end{equation}
\noindent
\begin{equation}
P_{BISEP}=\frac{1}{\sqrt{2}}(e^{123}+e^{156}),\qquad
\mathcal{D}(P)=0,\quad \tilde{P}=0 \label{3v}
\end{equation}
\noindent
\begin{equation}
P_{SEP}=e^{123},\qquad \mathcal{D}(P)=0,\quad \tilde{P}=0.
\label{4v}
\end{equation}
Here we have given the representatives of the GHZ and W classes in
a form different from the ones appearing in TABLE \ref{tab:1}. In this new
form the number of terms appearing in the representatives is
increasing as we proceed from the separable class to the maximally
entangled GHZ one. Notice that the difference from the
representatives of the GHZ and W classes of TABLE \ref{tab:1}. and
Eqs.(\ref{1v}) and (\ref{2v}) amounts to a SLOCC transformation.
The meaning of these transformations can easily be clarified if we
reinterpret these states as three-qubit ones according to the
prescription of Eq.(\ref{specampl}). Indeed, using the new
labelling of Eq.(\ref{atnevezes}) the three-qubit states
correponding to the ones of Eqs.(\ref{1v})-(\ref{4v}) are
\begin{equation}
\frac{1}{2}(\vert 000\rangle+\vert 011\rangle +\vert 101\rangle
+\vert 110\rangle) \label{1vv}
\end{equation}
\begin{equation}
\frac{1}{\sqrt{3}}(\vert 000\rangle +\vert 011\rangle +\vert
101\rangle) \label{2vv}
\end{equation}
\begin{equation}
\frac{1}{2}(\vert 000\rangle +\vert 011\rangle) \label{3vv}
\end{equation}
\begin{equation}
\vert 000\rangle. \label{4vv}
\end{equation}
\noindent Now it is easy to show that
\begin{equation}
\frac{1}{2}(\vert 000\rangle +\vert 011\rangle +\vert 101\rangle
+\vert 110\rangle)=(H\otimes H\otimes H)\frac{1}{\sqrt{2}}(\vert
000\rangle +\vert 111\rangle)=H\otimes H\otimes H\vert GHZ\rangle,
\label{GHZv2}
\end{equation}
\noindent and
\begin{equation}
\frac{1}{\sqrt{3}}(\vert 000\rangle +\vert 011\rangle +\vert
101\rangle)=(I\otimes I\otimes X)\frac{1}{\sqrt{3}}(\vert
001\rangle+\vert 010\rangle +\vert 100\rangle)=(I\otimes I\otimes
X)\vert W\rangle \label{Wv2}
\end{equation}
\noindent
 where $H$ and $X$ are  the usual
Hadamard and bit flip gates
\begin{equation}
H=\frac{1}{\sqrt{2}}\begin{pmatrix}1&1\\1&-1\end{pmatrix},\qquad
X=\begin{pmatrix}0&1\\1&0\end{pmatrix}. \label{gates}
\end{equation}
Hence, these states are local unitary (hence also SLOCC)
equivalent to the usual $GHZ$ and $W$ states\cite{Dur}. Notice
also that since
\begin{equation}
D((g_1\otimes g_2\otimes g_3)\psi)={\rm Det}(g_1)^2{\rm
Det}^2(g_2){\rm Det}^2(g_3)D(\psi),\qquad g_1,g_2,g_3\in GL(2,{\bf
C})\label{trafocayley}
\end{equation}
\noindent none of these transformations changes the value of
Cayley's hyperdeterminant $D(\psi)$. Moreover, since
$\mathcal{D}(P_{\psi})=D(\psi)$ after reinterpreting again our
three-qubit states as three-fermionic ones via the correpondence
$\psi\mapsto P_{\psi}$ the SLOCC transformations acting on the
corresponding fermionic states can be obtained from the ones of
Eq.(\ref{GHZv2})-(\ref{Wv2}) using
Eq.(\ref{szokasoshatas})-(\ref{szokasosbeagy}) and the permutation $(3245)$.

Notice also that in order to separate the last two classes with
representatives of Eq.(\ref{3v})-(\ref{4v}) one has to use the
Pl\"ucker relations of Eqs.(\ref{Penrose})-(\ref{general}). In our
special case these relations can be described in the following
elegant form\cite{Leclerc} \beq \eta X=Y^{\sharp},\qquad \xi
Y=X^{\sharp},\qquad \eta\xi I=XY \label{pluckerclerc} \eeq
\noindent where for the connection between the amplitudes of $P$
and the quantities $(\eta,X,Y,\xi)$ see
Eqs.(\ref{etaxi})-(\ref{Xmatr}). These relations hold if and only
if the corresponding fermionic state is separable, i.e. can be
written in the form of a single Slater determinant.

 The GHZ and W classes are the
two inequivalent classes for tripartite entangled fermionic
systems with six modes. These classes are completely characterized
by the relative invariant ${\cal D}(P)$ and the dual state
$\tilde{P}$ (a covariant). Note that the GHZ class corresponds to
a stable SLOCC orbit\cite{Satokimura}. Stability means that states
in a neighborhood (with respect to the Zariski topology) of a
particular one are all SLOCC equivalent ones. More precisely
states of the GHZ class form an open dense orbit within the state
space of three fermions with six single particle states.
 This fact is related
to the result that our state space of such fermions corresponds to
a prehomogeneous vector space which is the class No.5. in the
Sato-Kimura classification\cite{Satokimura} of such spaces.

 Let us elaborate on this stable class of GHZ states.
 As we know the canonical form of a representative from the genuine entangled (GHZ) class is \beq
P_0=e^{123} + e^{456}. \eeq For this representative , one can
easily check that the matrix of $K_{P_0}^{[1,1]}$ is of the form
\beq (K_{P_0}^{[1,1]})^a_{\;\;b}= \left(
\begin{array}{cccccc}
 1 & & & & & \\
 & 1 & & & & \\
 & & 1 & & & \\
 & & & -1 & & \\
 & & & & -1 & \\
 & & & & & -1
\end{array}
\right). \eeq
 Now
$\mathcal{D}(P_0)=1$ hence for the dual state of
Eq.(\ref{freudual}) we have \beq
\hat{P}_0=-i(e^{123}-e^{456}).\label{freudualghz} \eeq \noindent
Now the states $P_0+i\hat{P}_0$ and $P_0-i\hat{P}_0$ are clearly
separable ones. Moreover, since $P$ and $\hat{P}$ both transform
covariantly under SLOCC transformations  separability is preserved
hence for {\it any} state $P$ with $\mathcal{D}(P)\neq 0$ (i.e. a
one in the GHZ class) the states \beq U_{\pm}=P\pm
i\hat{P}\label{ghzkomponensek} \eeq \noindent are separable ones.
In other words for any state in the GHZ class the expression \beq
P=\frac{1}{2}(U_++U_-) \label{dekompozicio}\eeq \noindent provides
a canonical decomposition in terms of two Slater determinants.

Let us also discuss the structure of the SLOCC classes for {\it
real states}. In this case the vector space underlying our
three-fermion state space is $V={\mathbb R}^6$ and the SLOCC group
is $GL(6,{\mathbb R})$. In contrast to the five classes of TABLE
\ref{tab:1}. now we have {\it six entanglement classes}. The extra class is
coming from a splitting of the usual GHZ class into {\it two
classes}. The two classes are having $\mathcal{D}(P)>0$ and
$\mathcal{D}(P)<0$ are called $GHZ_+$ and $GHZ_-$ classes
respectively. The normalized representatives are \beq
P_{GHZ_+}=\frac{1}{2}(e^{123}+e^{156}+e^{264}+e^{345}),\qquad
P_{GHZ_-}=\frac{1}{2}(e^{123}-e^{156}-e^{264}-e^{345}).\label{ghzclassesreal}
\eeq \noindent Of course $P_{GHZ_+}$ is just the state known from
Eq.(\ref{1v}) which is real SLOCC equivalent to the GHZ
representative of TABLE \ref{tab:1}. Let us illustrate this result in the
language of embedded three-qubit systems. These fermionic states
correspond to the ones \beq \vert GHZ_+\rangle=\frac{1}{2}(\vert
000\rangle +\vert 011\rangle + \vert 101\rangle +\vert 110\rangle)
,\quad \vert GHZ_-\rangle=\frac{1}{2}(\vert 000\rangle -\vert
011\rangle - \vert 101\rangle -\vert 110\rangle).\label{realghzk}
 \eeq \noindent
We already know from Eq.(\ref{GHZv2}) that \beq \vert GHZ_+\rangle
=H\otimes H\otimes H\vert GHZ\rangle.\eeq \noindent On the other
hand we have \beq \vert GHZ_-\rangle
=\mathcal{U}\otimes\mathcal{U}\otimes \mathcal{U}\vert GHZ\rangle,
\qquad
\mathcal{U}=\frac{1}{\sqrt{2}}\begin{pmatrix}1&i\\1&-i\end{pmatrix}.
\label{ekvivalensghzk} \eeq \noindent These expressions illustrate
the fact that though the states $\vert GHZ_{\pm}\rangle$ are
complex SLOCC equivalent however, they are real SLOCC
inequivalent. One can also write these states as \beq \vert
GHZ_+\rangle =\frac{1}{\sqrt{2}}(\vert F_+\rangle\otimes \vert
F_+\rangle \otimes \vert F_+ \rangle+\vert F_-\rangle \otimes
\vert F_-\rangle \otimes \vert F_-\rangle),\qquad \vert
F_{\pm}\rangle= \frac{1}{\sqrt{2}}(\vert 0\rangle \pm\vert
1\rangle)\label{atvaltas1} \eeq \noindent
 \beq \vert GHZ_-\rangle
=\frac{1}{\sqrt{2}}(\vert E\rangle\otimes \vert E\rangle \otimes
\vert E \rangle+\overline{\vert E\rangle \otimes \vert E\rangle
\otimes \vert E\rangle}),\qquad \vert E\rangle=
\frac{1}{\sqrt{2}}(\vert 0\rangle + i\vert
1\rangle)\label{atvaltas2} \eeq \noindent where the overline means
complex conjugation. These expressions illustrate our result of
Eq.(\ref{dekompozicio}) for decomposing an arbitrary state from
the GHZ class into {\it two} separable states. According to the
definition of the dual state $\hat{P}$ of Eq.(\ref{freudual}) we
see that for real states  with $\mathcal{D}(P)>0$ the separable
components are remaining {\it real}, on the other hand for
$\mathcal{D}(P)<0$ they are {\it complex} conjugate states. In
this latter case one can define the $6\times 6$
matrix\cite{Hitchin} \beq J_P\equiv
\frac{K^{[1,1]}_P}{\sqrt{-\mathcal{D}(P)}} \eeq \noindent One can
prove that $(K^{[1,1]}_P)^2=\mathcal{D}(P) I$, where $I$ is the
$6\times 6$ identity matrix. Hence for $\mathcal{D}(P)<0$ we have
\beq J_P^2=-I.\label{komplexstruktura} \eeq\noindent This means
that if we start with a {\it real} three fermion state $P$
satisfying $\mathcal{D}(P)<0$ then on its single particle space
$V=\mathbb{R}^6$ this $P$ defines a {\it complex structure}. For
the special state $P_{GHZ_-}$ of Eq.(\ref{ghzclassesreal}) the
complex structure in question is just the canonical one giving
rise on $V$ to the complex coordinates 
\beq
E^{1,2,3}=e^{1,2,3}+ie^{4,5,6},\qquad
E^{\overline{1},\overline{2},\overline{3}}=e^{1,2,3}-ie^{4,5,6}.
\label{komplexcoord} 
\eeq 
\noindent Since \beq
E^{123}+E^{\overline{123}}=2(e^{123}-e^{156}-e^{426}-e^{345})\label{kakukkos}
\eeq \noindent in the three-qubit reinterpretation these complex
coordinates correspond to our writing $\vert GHZ_-\rangle$ in the
(\ref{atvaltas2}) form. Notice also that with respect to the
complex structure $J_{P_{GHZ_-}}$ the components $E^{123}$ and
$E^{\overline{123}}$ are giving rise to the $(3,0)$ holomorphic
and $(0,3)$ antiholomorphic parts of $P_{GHZ_-}$.

We note in closing that there is an interesting physical
application of these complex structures as defined by real
three-fermion states. For this one takes a closed oriented six
dimensional real manifold $\mathcal{M}$ equipped with a real
differential three-form $P$ with $\mathcal{D}(P)<0$ everywhere.
Notice that locally at each point of $\mathcal{M}$ the tangent
space and its dual gives rise to copies of a $V=\mathbb{R}^6$,
hence we can regard the differential three-form $P$ as a
collection of three-fermion states parametrized by the points of
$\mathcal{M}$. Now such a $P$ defines an {\it almost complex
structure} $J_P$ on $\mathcal{M}$. One can then show\cite{Hitchin}
that when $P$ is closed and belonging to a fixed cohomology class
then the critical points of the functional \beq
V_H=\int_{\mathcal{M}}\sqrt{-\mathcal{D}(P)}d^6x\label{hitchinfunct}
\eeq \noindent are satisfying the equation \beq
d\hat{P}=0\eeq\noindent meaning that the dual form $\hat{P}$ is
also closed. Hence the {\it separable} differential form
$\Omega=P+i\hat{P}$ is of type $(3,0)$, closed and the almost
complex structure $J_P$ is {\it integrable}. In this way one can
generate a complex structure to a six dimensional manifold
$\mathcal{M}$ rendering it to a {\it threefold}. Calabi-Yau
threefolds are particularly important actors in string theory.
Such spaces describe the structure of extra dimensions. The shapes
and volumes of such spaces are subject to quantum fluctuations. It
can be shown that the fluctuations in shapes preserving volume
correspond to fluctuations in the complex structure of
$\mathcal{M}$. Hence the result briefly discussed above means that
the critical points of certain action functionals of three-forms
belonging to a fixed cohomology class single out {\it special}
complex structures. Fixing a cohomology class physically means
that we fix the wrapping configurations of three-dimensional
extended objects , membranes, around the noncontractible
three-cycles of the extra dimensions. Under certain conditions the
projections of these higher dimensional configurations to our four
dimensional space-time look like charged black-holes. For an
application of these ideas within the interesting field of the so
called Black-Hole/Qubit Correspondence\cite{BHQC} see our recent
paper on Hitchin functionals related to measures of
entanglement\cite{levsar}.

\subsection{Seven single particle states}
\label{sec:threefseven}

In the case of three fermions with seven single particle states an
arbitary unnormalized state is described by the element
$\mathcal{P} \in\wedge^3V^*$ where $V=\mathbb{C}^7$. Such an
element can be written as \beq
\mathcal{P}=\frac{1}{3!}\mathcal{P}_{I_1I_2I_3}e^{I_1}\wedge
e^{I_2} \wedge e^{I_3} \eeq with $\lbrace e^I \rbrace_{I=1}^7$ a
basis of $V^*$. Now the SLOCC group is $GL(V)=GL(7,\mathbb{C})$
with the same kind of action as in \eqref{kformakomphatas}.

Let us first consider the covariants $\kappa_P^{(1)}\equiv \kappa_P^{[2,1]}$ and $\kappa_P^{(1,1)}\equiv \kappa_P^{[0,2]}$. For simplicity we introduce the notation:
\beq {(M^{A})^{B}}_C\equiv
{\left(K_P^{[2,1]}\right)^{AB}}_{C}\eeq\noindent \beq N_{AB}\equiv
\left(K_P^{[0,2]}\right)_{AB}
\eeq
\noindent where their explicit
form according to Eq.(\ref{nagykmatrixalakja}) is 
\beq
{(M^{A})^{B}}_C=\frac{1}{12}\varepsilon^{ABI_1I_2I_3I_4I_5}\mathcal{P}_{CI_1I_2}\mathcal{P}_{I_3I_4I_5}
\label{emformula} 
\eeq \noindent 
\beq
N_{AB}=\frac{1}{24}\varepsilon^{I_1I_2I_3I_4I_5I_6I_7}\mathcal{P}_{AI_1I_2}\mathcal{P}_{BI_3I_4}\mathcal{P}_{I_5I_6I_7}.
\label{enformula}
\eeq
\noindent Note that for later use we have
regarded $M$ as a collection of seven $7\times 7$ matrices. Also
note that in the real case $V={\mathbb R}^7$ used in the
literature on manifolds of special
holonomy\cite{Dijkgraaf,Hitchin,Bryant} a suitable scalar multiple
of the latter covariant shows up as\beq
\mathcal{B}_{AB}=-\frac{1}{6}N_{AB}.\label{Bmatrix}\eeq\noindent
It is arising from the map $\mathcal{B_P}:V\otimes V\rightarrow
\wedge^7V^{\ast}$ which gives rise to a seven form when acting on
the pair of vectors $v$ and $u$ as 
\beq
\mathcal{B_P}(v,u)=-\frac{1}{6}\iota_v\mathcal{P}\wedge\iota_u\mathcal{P}\wedge
\mathcal{P}.
\eeq\noindent 
The transformation properties of these covariants
are 
\beq
 {(M^A)^B}_C\mapsto ({\rm Det}g^{\prime})
{g^A}_D{g^B}_E{{g^{\prime}}_C}^F {(M^D)^E}_F
\label{emtrafozik}
\eeq \noindent
 \beq
  N_{AB}\mapsto ({\rm
Det}g^{\prime}) {{g^{\prime}}_A}^C
{{g^{\prime}}_B}^DN_{CD}.\label{entrafozik} 
\eeq \noindent

It is convenient to study the case of three fermions with seven
single particle states as the case of adding an extra mode to the
six original ones of three fermions discussed in the previous
section. For this purpose we split our seven dimensional vector
space $V$ to the direct sum of a six and a one dimensional vector
space spanned by the extra basis vector $e^7$. Then we write \beq
\mathcal{P}=P+\omega\wedge e^7 \label{fontossplit} \eeq \noindent
where $P$ is given by Eq.(\ref{3fw6}) and $\omega$ is a two-form
\beq \omega=\frac{1}{2}\omega_{ij}e^{i}\wedge
e^{j}.\label{twoformomega} \eeq \noindent In the following we
adopt the convention for the indices like $A,B,\dots I,J,\dots$
running from $1$ to $7$ on the other hand indices like $a,b,\dots
i,j,\dots$ are running from $1$ to $6$. Hence we have \beq
\mathcal{P}_{abc}\equiv P_{abc},\qquad
\mathcal{P}_{ab7}=\omega_{ab}.\label{rangekonv} \eeq \noindent

Now for the components of our covariants a straightforward
calculation yields the following results
 \beq {(M^7)^7}_7=0,\qquad {(M
^7)^7}_c=0,\qquad
{(M^7)^b}_7=\frac{1}{12}\varepsilon^{bijklm}\omega_{ij}P_{klm},\qquad
{(M^7)^b}_c={K^b}_c\label{mesek1}\eeq\noindent \beq
{(M^a)^7}_7=-\frac{1}{12}\varepsilon^{aijklm}\omega_{ij}P_{klm},\qquad
{(M^a)^7}_c=-{K^a}_c, \qquad
{(M^a)^b}_7=\frac{1}{4}\varepsilon^{abijkl}\omega_{ij}\omega_{kl}
\label{mesek2} \eeq\noindent \beq
{(M^a)^b}_c=\frac{1}{4}\varepsilon^{abijkl}(P_{cij}\omega_{kl}-\frac{2}{3}\omega_{ci}P_{jkl}).
\label{mesek3} \eeq \noindent \beq N_{77}=6{\rm Pf}(\omega),\qquad
N_{a7}=N_{7a}=-\frac{1}{12}\varepsilon^{ijklmn}(\omega_{ai}\omega_{jk}P_{lmn}+\frac{2}{3}P_{aij}\omega_{kl}\omega_{mn}).
\label{nesek1} \eeq \noindent \beq
N_{ab}=\frac{1}{8}\varepsilon^{ijklmn}P_{aij}P_{bkl}\omega_{mn}+{K^c}_a\omega_{cb}+{K^c}_b\omega_{ac}.
\label{nesek2}\eeq\noindent Here by an abuse of notation for the
covariant of Eq.(\ref{Kmatrixamikell}) we have used the shorthand
${K^a}_b$ and ${\rm Pf}(\omega)$ is the Pfaffian of $\omega$
defined as \beq {\rm
Pf}(\omega)=\frac{1}{2^33!}\varepsilon^{ijklmn}\omega_{ij}\omega_{kl}\omega_{mn}.\label{pfaffian}
\eeq \noindent

These expressions can be further simplified in the special case
when \beq P\wedge \omega=0.\label{konditionprimitive} \eeq
\noindent The meaning of this condition is as follows. Due to the
split of Eq.(\ref{fontossplit}) one can understand the structure
of three-fermions with seven single particle states with $35$
amplitudes via looking at the simpler structure of three fermions
with six single particle states having merely $20$ ones. In this
perspective the two-form $\omega$ giving rise to the extra $15$
amplitudes can be regarded as an extra structure living on the six
dimensional vector space: a symplectic form. The condition
$P\wedge \omega=0$ encapsulates a compatibility condition between
the symplectic form $\omega$ and the three-form $P$. In the
mathematical literature this condition means that the three-form
$P$ is {\it primitive} with respect to $\omega$. When $P$ is
primitive
 using the identity
$\iota_{e^i}(P\wedge\omega)=\iota_{e^i}P\wedge
\omega-P\wedge\iota_{e^i}\omega$ one can show that \beq
\left(P_{aij}\omega_{kl}+\frac{2}{3}\omega_{ai}P_{jkl}\right)e^i\wedge
e^j\wedge e^k\wedge e^l=0.\label{explicitprimitiv}\eeq\noindent
For our covariants this result yields the much simpler looking
expressions
 \beq {(M^7)^7}_7={(M
^7)^7}_c={(M^7)^b}_7=0,\qquad
{(M^7)^b}_c={K^b}_c\label{mesek1v}\eeq\noindent \beq
{(M^a)^7}_7=0,\quad {(M^a)^7}_c=-{K^a}_c, \quad
{(M^a)^b}_7=\frac{1}{4}\varepsilon^{abijkl}\omega_{ij}\omega_{kl},\quad
{(M^a)^b}_c=\frac{1}{2}\varepsilon^{abijkl}P_{cij}\omega_{kl}
\label{mesek2v}\eeq \noindent \beq N_{77}=6{\rm Pf}(\omega),\qquad
N_{a7}=N_{7a}=0,\qquad
N_{ab}=3{K^c}_a\omega_{cb}=3{K^c}_b\omega_{ca}.
\label{nesek2v}\eeq\noindent Notice that by virtue of
Eq.(\ref{nesek2v}) the $7\times 7$ matrix $\boldsymbol{N}$ can be
written in the factorized form \beq \boldsymbol{N}= \left(
\begin{array}{c|c}
-3\boldsymbol{\omega} & 0 \\
\hline 0 & 6\text{Pf}(\boldsymbol{\omega})
\end{array}
\right) \left(
\begin{array}{c|c}
\boldsymbol{K}& 0 \\
\hline 0 & 1
\end{array}
\right)\label{enfaktorosan}\eeq where $\boldsymbol{\omega}$  and
$\boldsymbol{K}$ are the $6\times 6$ matrices corresponding to the
coefficient matrix of the two form $\omega$ and the matrix of
Eq.(\ref{Kmatrixamikell}). We see that if the first matrix has
full rank then rank $\boldsymbol{N}=$rank$\boldsymbol{K}+1$. Also
because the matrix $\boldsymbol{N}$ has to be symmetric,
$\boldsymbol{\omega}$ and $\boldsymbol{K}$ must anticommute.

It is convenient to define another symmetric $7\times 7$ matrix
\beq L^{AB}\equiv {(M^A)^C}_D{(M^B)^D}_C.\label{dualismetrika}
\eeq \noindent This matrix is a covariant quartic in the original
amplitudes and under SLOCC transformations transforms as \beq
L^{AB}\mapsto ({\rm
Det}g^{\prime})^2{g^A}_C{g^B}_DL^{CD}.\label{elinvtrafo} \eeq
\noindent

The $7\times 7$ matrix $\boldsymbol{L}$ can be regarded as the
dual of the one $\boldsymbol{N}$. In Appendix A. it is shown that
for $P\wedge \omega=0$ this matrix can also be written in the
factorized form \beq \boldsymbol{L}= \left(
\begin{array}{c|c}
-12\boldsymbol{\tilde{\omega}} & 0 \\
\hline 0 & 6
\end{array}
\right) \left(
\begin{array}{c|c}
\boldsymbol{K}& 0 \\
\hline 0 & \mathcal{D}(P)
\end{array}
\right)\label{elfaktorosan}\eeq where \beq
\tilde{\omega}^{ij}=\frac{1}{8}\varepsilon^{ijklmn}\omega_{kl}\omega_{mn}.\label{omegatilde}
\eeq \noindent

From the covariants $\boldsymbol{N}$ and $\boldsymbol{L}$ one can
form a relative invariant $\mathcal{J}(\mathcal{P})$ homogeneous
of degree $7$ \beq \mathcal{J}(\mathcal{P})\equiv \frac{1}{2^4
3^27}{\rm
Tr}(\boldsymbol{LN})=\frac{1}{2^43^27}L^{AB}N_{AB}\label{sevenrelinv}
\eeq \noindent where the normalization was chosen for future
convenience. Under SLOCC transformations we have \beq
\mathcal{J}(\mathcal{P})\mapsto ({\rm
Det}g^{\prime})^3\mathcal{J}(\mathcal{P}).\label{jetrafoja} \eeq
\noindent Clearly when written in terms of the components of $P$
and $\omega$ the relative invariant $\mathcal{J}(\mathcal{P})$ has
a complicated expression. However, by virtue of
Eqs.(\ref{enfaktorosan}) and (\ref{elfaktorosan}) in the special
case when $P\wedge\omega=0$ it has a factorized form \beq
\mathcal{J}(\mathcal{P})=\frac{1}{4}{\rm
Pf}(\omega)\mathcal{D}(P).\label{invariansfaktorizalt} \eeq
\noindent

As a useful relative invariant one can also define either ${\rm
Det}(\boldsymbol{N})$ or ${\rm Det}(\boldsymbol{L})$. However, it
is easy to see that none of them is independent from
$\mathcal{J}(\mathcal{P})$. Indeed, using e.g.
Eq.(\ref{enfaktorosan}) for calculating ${\rm
Det}(\boldsymbol{N})$ one obtains \beq {\rm
Det}(\boldsymbol{N})=-6\cdot\left(9{\rm
Pf}(\omega)\mathcal{D}(P)\right)^3.\label{determinansrelinv} \eeq
\noindent Here we have used ${\rm Det}(K)=-\mathcal{D}^3$ which
follows from $K^2=\mathcal{D}\boldsymbol{1}$ , ${\rm Tr}(K)=0$ and
the Newton identities, moreover we have also used that ${\rm
Det}(\omega)=\left({\rm Pf}(\omega)\right)^2$. The case of real states is important in the string theory literature where the determinant of the matrix $\mathcal{B}_{\mathcal{P}}$ of
Eq.(\ref{Bmatrix}) is used\cite{Dijkgraaf} as a relative invariant. With our normalization
as used in Eq.(\ref{sevenrelinv}) we have 
\beq {\rm
Det}\mathcal{B}_{\mathcal{P}}=\left(\mathcal{J}(\mathcal{P})\right)^3.
\label{kobososszefugges}
\eeq 
\noindent

\begin{figure}[h!]
\includegraphics[width=0.4\textwidth]{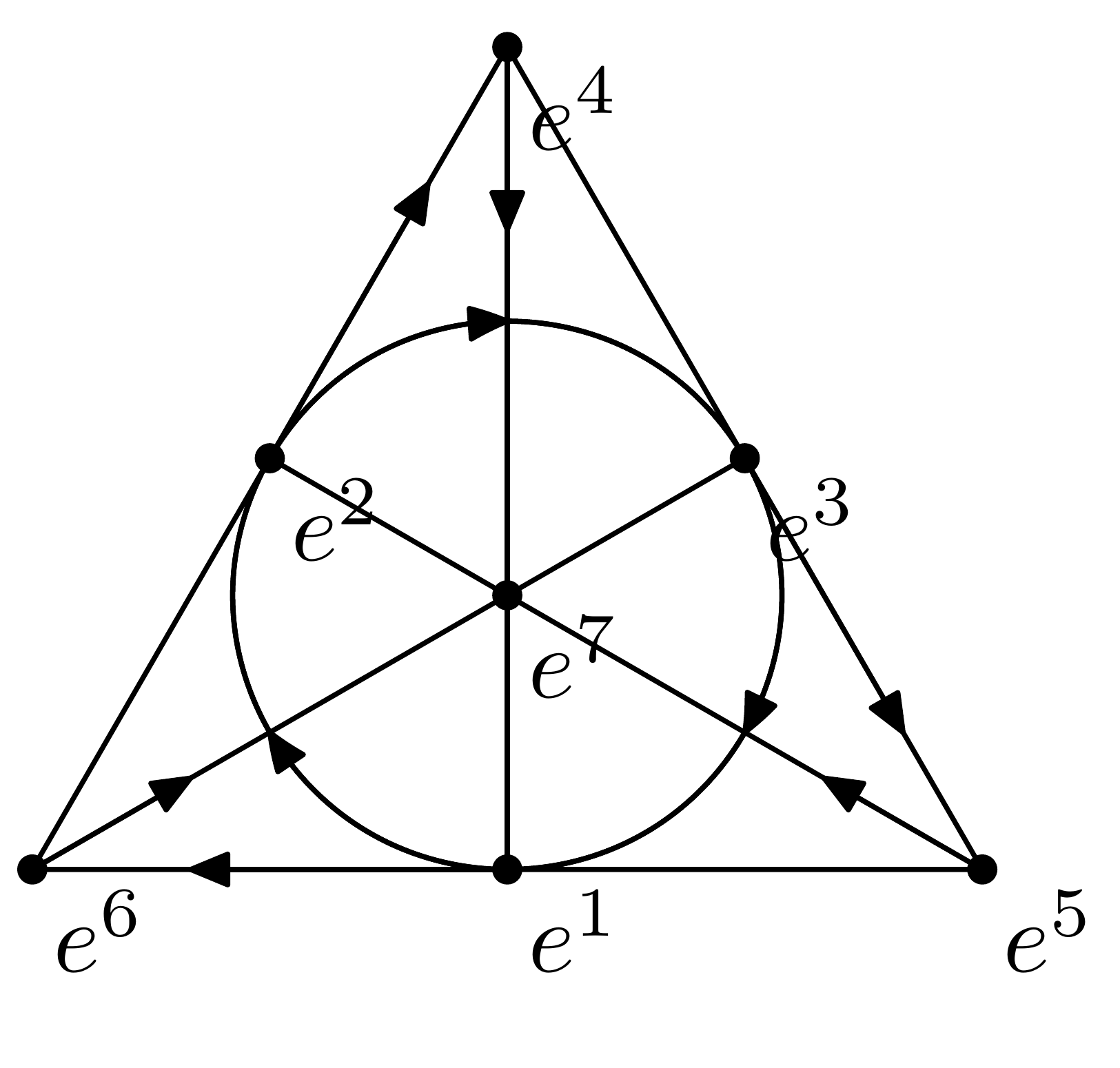}
\caption{The oriented Fano plane. The points of the plane correspond to the basis vectors of the seven dimensional single particle space. The lines of the plane represent three fermion basis vectors with the arrows indicating the order of single particle states in them to get a plus sign.}
\label{fig:1}
\end{figure}

In order to present the SLOCC classification of three fermions
with seven single particle states let us consider again our seven
dimensional complex vector space $V$ with its canonical basis
vectors by $e_A$. Let us denote as usual the basis vectors of its
six dimensional subspace by $e_a, a=1,\dots 6$. As a complex basis
of the dual of this subspace we define \beq
E^{1,2,3}=e^{1,2,3}+ie^{4,5,6},\qquad
E^{\overline{1},\overline{2},\overline{3}}=e^{1,2,3}-ie^{4,5,6},
\qquad E^7=ie^7. \label{ebasis} \eeq \noindent Then a $GHZ$-like
state in the six dimensional subspace can be written as \beq
E^{123} +E^{\overline{1}\overline{2}
\overline{3}}=2(e^{123}-e^{156} +e^{246}-e^{345})
\label{naittenvanazghz}\eeq \noindent With the usual relabelling
$4, 5, 6\mapsto \overline{1},\overline{2},\overline{3}$ and up to
normalization the state on the right hand side is just the one of
Eq.(\ref{kakukkos}).
 Let us add to this state the one $(E^{1\overline{1}}+ E^{2
\overline{2}}+E^{3\overline{3}})\wedge E^7$. This contains a full
rank symplectic form of that six dimensional subspace in complex
form. Then as our basic three fermion state with seven single
particle states we chose

\beq 
\begin{aligned}
{\mathcal{P}}_0 &\equiv
\frac{1}{2}(E^{123}+E^{\overline{123}}+(E^{1\overline{1}}
+E^{2\overline{2}}+E^{3\overline{3}})\wedge
E^7)\\
&=e^{123}-e^{156}+e^{246}-e^{345} +e^{147}+e^{257}+e^{367}.
\end{aligned}
\label{calibration} 
\eeq \noindent Notice that the structure of
our tripartite state ${\mathcal{P}}_0$ is encoded in the incidence
structure of the {\it lines} of the {\it oriented} Fano plane
which is also encoding the multiplication table of the octonions (see FIG. \ref{fig:1}.).
As a {\it complex} three-form it can be
shown\cite{Schouten,Bryant} that the subgroup ${\rm
Stab}(\mathcal{P}_0)$ of the SLOCC group $GL(7,{\mathbb C})$ that
fixes ${\mathcal{P}}_0$ is the exceptional group $G^{\mathbb
C}_2\times\{\omega{\bf 1}\vert\omega^3=1\}$ where ${\bf 1}$ is the
$7\times 7$ identity matrix. From the theory of prehomogeneous
vector spaces\cite{Satokimura,Kimura} it is known that the
three-form $\mathcal{P}_0$ can be regarded as the representative
of the Zariski-open SLOCC orbit of the prehomogeneous vector space
$(\wedge^3V,GL(7,{\mathbb C}),\varrho)$. Here $\varrho$ refers to
the representation of $G\equiv GL(7,{\mathbb C})$ on $W\equiv
\wedge^3V$ of the (\ref{kformakomphatas}) form i.e. the one
induced by the canonical representation of $G$ on $V$. The orbit
determined by $\mathcal{P}_0$ is dense, meaning its Zariski
closure gives the full space $\wedge^3V$. Notice that ${\rm
dim}\wedge^3V=35$, ${\rm dim}G=49$ and $ {\rm dim}{\rm
Stab}(\mathcal{P}_0)={\rm dim}G_2=14$ hence ${\rm dim}G={\rm
dim}W+{\rm dim}{\rm Stab}(\mathcal{P}_0)$.

A comment here is in order. Rather than using $\mathcal{P}_0$ as
an entangled state, in string theory it is used as a {\it real}
differential form on a seven dimensional real {\it manifold}. In
this context instead of the complex SLOCC group the real one i.e.
$GL(7,{\mathbb R})$ is used. The stabilizer of ${\mathcal{P}}_0$
as a real three-form is the compact real form $G_2$ which is the
automorphism group of the octonions. In the theory of special
holonomy manifolds invariant forms like $\mathcal{P}_0$ are called
calibrations. Note that after the permutation
$e^5\leftrightarrow e^7$ we obtain the form for $\mathcal{P}_0$
usually used in the literature on such
manifolds\cite{Becker,Bryant}.

\begin{figure}[h!]
\includegraphics[width=0.9\textwidth]{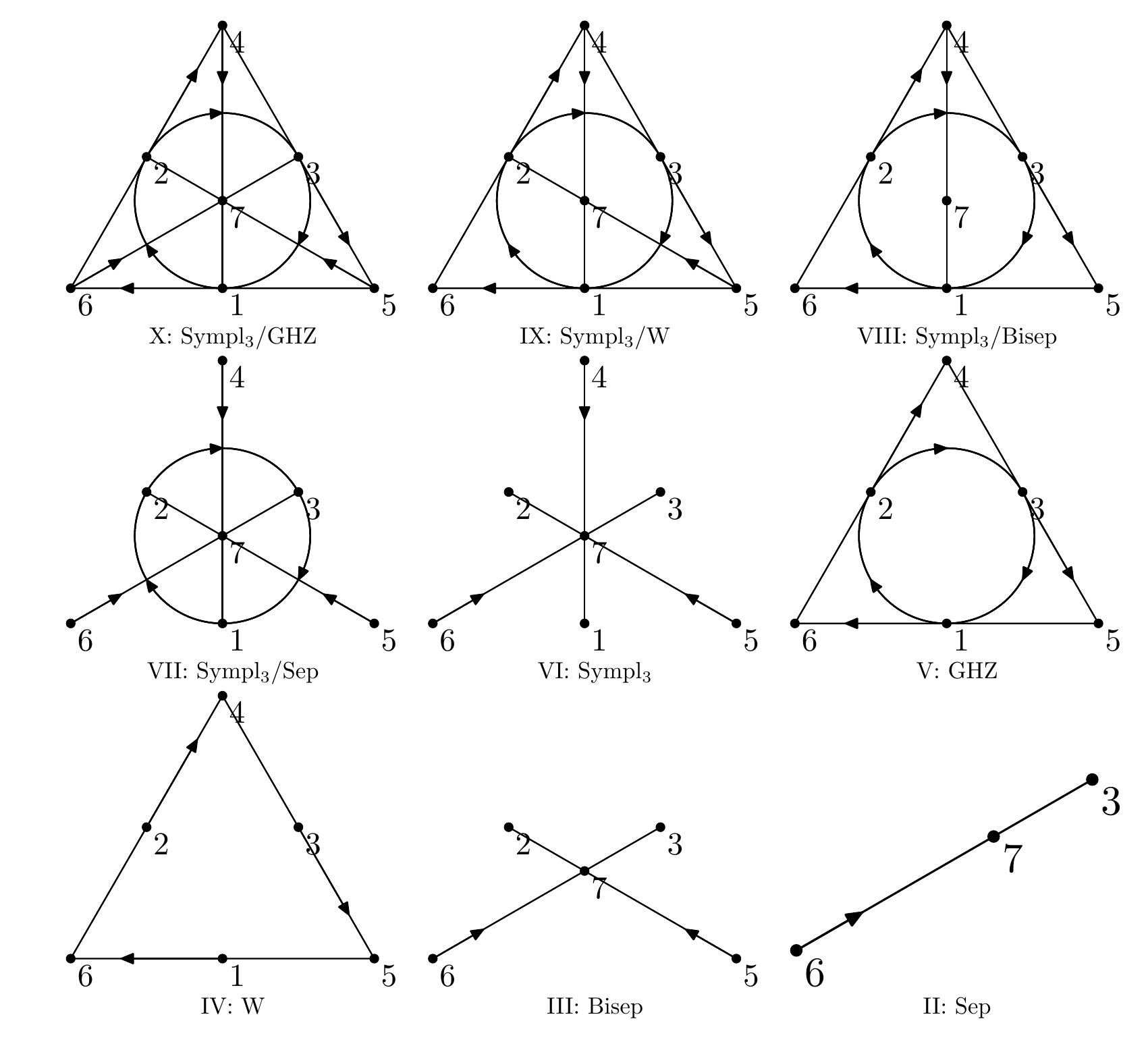}
\caption{Graphical representation of the nine entanglement classes of three fermions with seven single particle states with the use of the Fano plane.}
\label{fig:2}
\end{figure}

 The orbit structure of three fermions with seven single particle states is
available in the mathematical literature\cite{Reichel,Gurevich} and summarized in TABLE \ref{tab:2}. with the ranks of the basic covariants computed.
Here we would like to point out an important fact not mentioned in
the literature that the structure of these SLOCC classes can
elegantly be described using the Fano plane as follows (see FIG. \ref{fig:2}.).

{\bf Class I. (NULL.)} This is the null class consisting of the
trivial zero state.

{\bf Class II. (SEP.)} This is the class of separable states
consisting of a single Slater determinant. As a representative of
this class we chose the $e^{367}$ part of the state
$\mathcal{P}_0$ of Eq.(\ref{calibration}). We can chose a
graphical representation for this state as an oriented circle with
three distinguished points $3$, $6$ and $7$. Alternatively, after
remembering that the numbers $367$ can be cyclically permuted
without introducing a sign, we can represent this state as an
oriented line $(673)$ of the Fano plane starting from the point
$6$ and ending at the point $3$.

{\bf Class III. (BISEP.)} This is the class of biseparable states.
As a representative of this class we chose the $e^{257}+e^{367}$
part of the state $\mathcal{P}_0$. Graphically we can refer to
this class as two oriented circles touching each other at the
point $7$. Alternatively, using a cyclic rearrangement, one can
depict this class as the two oriented lines $572$ and $673$ of the
Fano plane intersecting at the point $7$.

{\bf Class IV. (W.)} This is the SLOCC class of W-states with
representative taken to be the $e^{246}-e^{345}-e^{156}$ part of
$\mathcal{P}_0$. Notice that in the notation of
Eq.(\ref{atnevezes}) using a cyclic rearrangement this state is of
the $-e^{\dot{1}2\dot{3}}-e^{\dot{1}\dot{2}3}-e^{1\dot{2}\dot{3}}$
form which is the negative of a state reminiscent of the
three-qubit W-state hence the name. For a graphical representation
one can imagine three oriented circles touching each other in the
points $4$ , $5$ and $6$. Alternatively one can take the three
oriented lines $(435)$, $(516)$ and $(624)$ of the Fano plane
forming a clock-wise oriented triangle with its vertices taken as
the points $4$, $5$ and $6$.

{\bf Class V.(GHZ.)} This is the SLOCC class of GHZ-states with
representative taken to be the $e^{123}-e^{156}+e^{246}-e^{345}$
part of the state $\mathcal{P}_0$. To see that this state is SLOCC
equivalent to the usual two-term GHZ-state just refer to
Eq.(\ref{naittenvanazghz}). For a graphical representation one can
envisage the equilateral triangle of Class IV. with a clock-wise
oriented circle inserted in the middle touching the triangle in
the three points : $1$, $2$ and $3$. Clearly the resulting picture
is just that part of the Fano plane that we obtain after omitting
the three lines intersecting in the point $7$.

{\bf Class VI. (SYMPL3/NULL.)} This class is a one whose
representative is based on a symplectic form with rank $3$ defined
on a six dimensional subspace of $V$. The representative we chose
is just the one based on the symplectic form
$e^{14}+e^{25}+e^{36}$. This gives rise to the
 $e^{147}+e^{257}+e^{367}$ part of $\mathcal{P}_0$.
 As a graphical representation for this class we chose the three
 lines $572$, $673$ and $471$of the Fano plane intersecting in the point
 $7$.
Notice that the arising diagram is just the complement of that
part of the Fano plane which represents the GHZ-class.

{\bf Class VII.(SYMPL3/SEP.)} This class is represented by a full
rank symplectic form plus a separable state. The corresponding
representative is the $e^{123}+e^{147}+e^{257}+e^{367}$ part of
$\mathcal{P}_0$. The graphical picture we can attach to this case
is that part of the Fano plane which consists of three lines
intersecting in the point $7$ and a circle $123$. This picture is
just the complement of the triangle part of the Fano plane
corresponding to the $W$-class.

{\bf Class VIII.(SYMPL1/GHZ.)} This class is represented by the
four-term GHZ-state plus a term containing a rank one part from
the symplectic form. This means that we keep the following five
terms from $\mathcal{P}_0$:
$e^{123}-e^{156}+e^{246}-e^{345}+e^{147}$. The resulting diagram
is containing five lines of the Fano plane. These are the ones
that are the complements of the two lines that show up in the
class of biseparable states.

{\bf Class IX. (SYMPL2/GHZ.)} This class is represented by the
four-term GHZ-state plus two terms containing the rank two part of
the symplectic form. In this case we keep the following six terms
from $\mathcal{P}_0$:
$e^{123}-e^{156}+e^{246}-e^{345}+e^{147}+e^{257}$. The
corresponding diagram contains six lines of the Fano plane. These
are the ones that form the complement of the single line $(673)$
showing up in the separable class.

{\bf Class X. (SYMPL3/GHZ.)} This is the class which corresponds
to the Zariski dense orbit in the space of three-forms. It is
represented by the state $P_0$ itself. Clearly since now we keep
all seven Slater determinants: the four ones comprising the
GHZ-state and the three ones giving rise to the full rank
symplectic form. The graphical representation of this class is
just the Fano plane itself. For the sake of completeness we
mention that this case can be regarded as the complement of the
null class represented by the zero state.

Using this graphical representation based on the Fano plane one
can obtain an alternative description of the SLOCC classes. First
observe that one can organize the ten classes into dual
pairs\cite{Gurevich}. These pairs are as follows $(I,X)$,
$(II,IX)$, $(III,VIII)$, $(IV,VII)$, $(V,VI)$. The pairs exhibit
complementary sets of lines of the Fano plane. The five dual pairs
can be labelled by the classes I-V that are just the well-known
five classes of three-fermions with six single particle states.
Some of the remaining five classes, namely classes VI, VII and X
can be labelled by a full rank symplectic form (SYMPL3) plus
representatives from the classes I-V (NULL,SEP,GHZ). However,
using the finite geometry of the Fano plane one can easily see
that even the exceptional classes i.e. VIII and IX can be given
this interpretation based on a full rank symplectic form. Indeed,
class VIII which is SYMPL1/GHZ can be reinterpreted as
SYMPL3/BISEP, on the other hand class IX which is SYMPL2/GHZ can
be reinterpreted as SYMPL3/W. In order to see this just look at
the diagram representing class IX. with a representative state
having six Slater determinants. Take the triple of lines $(624)$,
$(354)$ and $(714)$. Taken together they form a three-term state:
$(e^{62}+e^{35}+e^{71})\wedge e^{4}$ which is based on a the full
rank symplectic form $e^{62}+e^{35}+e^{71}$ in the six dimensional
subspace spanned by $\{e^1,e^2,e^3,e^5,e^6,e^7\}$. Take now the
remaining three oriented lines $(231)$, $(165)$ and $(572)$. It is
easy to see that in our new six dimensional subspace the
corresponding states form a $W$-state. Indeed, the oriented
triangle graphically representing such a W-state has now vertices
the points $1$, $5$ and $2$. Taking the permutation $(16)(2473)$
this new W-state and symplectic form is transformed back to the
one familiar from Classes IV. and VI. Similar reasoning gives the
desired reinterpretation for the class VIII. Now in this new
interpretation apart from the presence of a full rank symplectic
form the extra five classes VI-X are having the same structure as
the classes I-V. The upshot of these considerations is summarized
in TABLE \ref{tab:2}.

\begin{table}[h!]
\centering
\begin{tabular}{|c|c|c|c|c|c|}
\hline Name & Type & Canonical form of $\mathcal{P}$ & Rank
$\kappa_\mathcal{P}^{(1,1)}$ & Rank $\mathcal{P}^{(2)}$ & Rank
$\kappa_\mathcal{P}^{(1)}$\\ \hline \hline
I & NULL & 0 & 0 & 0 & 0\\
II & SEP & $E^{123}$ & 0 & 3 & 0\\
III & BISEP & $E^1\wedge (E^{23}+E^{\bar 2\bar 3})$ & 0 & 5 & 1\\
IV & W & $E^{12\bar 3}+E^{1\bar 2 3} + E^{\bar 1 2 3}$ & 0 & 6 & 3\\
V & GHZ & $E^{123} +E^{\bar 1\bar 2\bar 3}$ & 0 & 6 & 6\\
VI & SYMPL/NULL & $(E^{1\bar 1}+E^{2\bar 2}+E^{3\bar 3})\wedge E^7$ & 1 & 7 & 1\\
VII & SYMPL/SEP & $(E^{1\bar 1}+E^{2\bar 2}+E^{3\bar 3})\wedge E^7 +
E^{123}$ & 1 & 7 & 4\\
VIII & SYMPL/BISEP & $(E^{1\bar 1}+E^{2\bar 2}+E^{3\bar
3})\wedge E^7+E^1\wedge (E^{23}+E^{\bar 2\bar 3})$ & 2 & 7 & 6\\
IX & SYMPL/W & $(E^{1\bar 1}+E^{2\bar 2}+E^{3\bar 3})\wedge E^7+E^{12\bar 3}+E^{1\bar 2 3} + E^{\bar 1 2 3}$ & 4 & 7 & 7\\
X & SYMPL/GHZ & $(E^{1\bar 1}+E^{2\bar 2}+E^{3\bar
3})\wedge E^7+E^{123} +E^{\bar 1\bar 2\bar 3}$ & 7 & 7 & 7\\
\hline
\end{tabular}
\caption{Entanglement classes of three fermions with seven single particle states.}
\label{tab:2}
\end{table}

\eject

\subsection{Eight single particle states}

As usual define the three fermion state $P \in \wedge^3 V^*,\;
V=\mathbb{C}^8$ as 
\beq P=\frac{1}{3!}P_{i_1i_2i_3}e^{i_1}\wedge
e^{i_2} \wedge e^{i_3}, 
\eeq with $\lbrace e^i\rbrace_{i=1}^8$
being a basis of $V^*$. We have $GL(V)=GL(8,\mathbb{C})$ as the
SLOCC group with action identical as of Eq.(\ref{eq:sloccaction}).

According to Eqs.(\ref{megrondabb}) and (\ref{nagykmatrixalakja})
for three fermions with eight single particle states one can
define the covariants ${{\kappa}_P}^{[m,n]}$ with $3+2n+m=N=8$
with the corresponding matrix elements
${\left({K_P}^{[m,n]}\right)^{a_1\cdots a_m}}_{b_1\cdots b_n}$. We
need two such covariants based on 
\beq
\left(F^a\right)_{b_1b_2}\equiv{\left(K_P^{[1,2]}\right)^a}_{b_1b_2}
=\frac{1}{24}\varepsilon^{ai_1i_2i_3i_4i_5i_6i_7}P_{b_1i_1i_2}P_{b_2i_3i_4}P_{i_5i_6i_7}
\label{eftenzor}
\eeq\noindent 
and 
\beq
\left(E^{a_1a_2a_3}\right)_{b}\equiv{\left(K_P^{[3,1]}\right)^{a_1a_2a_3}}_{b}
=\frac{1}{12}\varepsilon^{a_1a_2a_3i_1i_2i_3i_4i_5}P_{bi_1i_2}P_{i_3i_4i_5}.
\label{hatenzor}
\eeq
\noindent

From one of these one can form the $8\times 8$ symmetric matrix
$\boldsymbol{G}$ which is of degree six in $P$ and transforming just as $N_{AB}$ of Eq.(\ref{entrafozik}):
\beq
G_{ab}\equiv\left(F^c\right)_{ad}\left(F^d\right)_{bc}.
\eeq\noindent
Alternatively one can define an $8\times 8$ symmetric matrix of
degree ten in $P$ as follows \beq
H^{ab}=\left(F^a\right)_{ci}\left(E^{ckl}\right)_j
\left(F^b\right)_{dk}\left(E^{dij}\right)_l. \label{dualisnyolc}
\eeq\noindent 
This quantity transform as \beq H^{ab}\mapsto ({\rm
Det}g^{\prime})^4{g^a}_c{g^b}_dH^{cd}.\label{naezistrafozik} \eeq
\noindent Now using the matrices $\boldsymbol{G}$ and
$\boldsymbol{H}$ one can form the relative invariant of degree
$16$ 
\beq
{\mathcal{I}}(P)={\rm
Tr}(\boldsymbol{GH})
\label{degreetizenhat} 
\eeq \noindent
transforming as \beq \mathcal{I}(P)\mapsto \left({\rm
Det}(g^{\prime})\right)^6\mathcal{I}(P).
\label{tizenhatosigytrafozik}\eeq\noindent

 The orbit structure of $\wedge^3 V^*$ is available in the
mathematical literature\cite{Gurevich}. It turns out that in
addition to the 10 classes of the previous section we have 13 more
classes. From the above and considerations of section
\ref{sec:covariants}, we have the non continuous independent invariants:
rank$G$, rank$F$=rank$\kappa_P^{(11)}$, rank$\kappa^{(1)}_P$ and rank$P^{(1)}$ so
far to classify these. It turns out that this is not sufficient
for full classification, we need to use the degree five map
$(F\bullet E)^{akl}_{ij}\equiv
\left(F^a\right)_{ci}\left(E^{ckl}\right)_j$. The rank of this map
is now sufficient for the full SLOCC classification.

\begin{table}[h!]
\centering
\begin{tabular}{|c|ccccccc|c|c|c|c|}
\hline Name & $\alpha$ & $\beta$ & $\gamma$ & $\delta$ &
$\epsilon$ & $\lambda$ & $\mu$ & Rank $G$ & Rank $F$ & Rank
$\kappa^{(1)}_P$ & Rank $F \bullet E$\\ \hline \hline
XI & 0 & 0 & 1 & 1 & 1 & 0 & 1  & 0 & 3 & 6 & 0\\
XII & 0 & 1 & 1 & 1 & 1 & 0 & 1  & 0 & 4 & 7 & 0\\
XIII & 1 & 1 & 1 & 0 & 0 & 0 & 1  & 0 & 4 & 8 & 0\\
XIV & 1 & 1 & 1 & 1 & 0 & 0 & 1  & 0 & 5 & 8 & 1\\
XV & 1 & 1 & 1 & 1 & 1 & 0 & 1  & 0 & 6 & 8 & 2\\
XVI & 0 & 0 & 1 & 0 & 0 & 1 & 1  & 1 & 8 & 8 & 1\\
XVII & 0 & 0 & 1 & 1 & 0 & 1 & 1  & 1 & 8 & 8 & 2\\
XVIII & 0 & 1 & 1 & 1 & 0 & 1 & 1  & 1 & 8 & 8 & 4\\
XIX & 0 & 0 & 1 & 1 & 1 & 0 & 1  & 2 & 8 & 8 & 2\\
XX & 0 & 1 & 1 & 1 & 1 & 1 & 1  & 2 & 8 & 8 & 5 \\
XXI & 1 & 1 & 1 & 0 & 0 & 1 & 1  & 3 & 8 & 8 & 7 \\
XXII & 1 & 1 & 1 & 1 & 0 & 1 & 1  & 5 & 8 & 8 & 8 \\
XXIII & 1 & 1 & 1 & 1 & 1 & 1 & 1  & 8 & 8 & 8 & 8 \\ \hline
\end{tabular}
\caption{Entanglement classes of three fermions with eight single particle states. The classes I,...,X of TABLE \ref{tab:2}. are omitted here.}
\label{tab:3}
\end{table}

The entanglement classes and the corresponding ranks are shown in TABLE \ref{tab:3}. The representative states are encoded as
\beq
\Lambda= \alpha E^{123} +\beta E^{567} +\gamma E^{154} +\delta E^{264} +\epsilon E^{374} +\lambda E^{278} +\mu E^{368},
\eeq
where as ususal $E^{ijk}=e^i\wedge e^j\wedge e^k$. The continuous invariant of Eq. \eqref{degreetizenhat} is only non-zero for the class XXIII which is a Zariski-open orbit of the prehomogeneous vector space\cite{Satokimura,Kimura} $(\wedge^3V,GL(8,\mathbb{C}),\varrho)$. Here again $\varrho$ is the representation \eqref{kformakomphatas} of $GL(8,\mathbb{C})$ on $\wedge^3 V$. 

\subsection{Nine single particle states}
\label{sec:threefnine}

Again, write a three fermion state as
\beq
P=\frac{1}{3!}P_{i_1i_2i_3}e^{i_1}\wedge e^{i_2} \wedge e^{i_3} \in \wedge^3 V^*,
\eeq
where $\lbrace e^i \rbrace_{i=1}^9$ is a basis of $V^*$ and now $V=\mathbb{C}^9$. The SLOCC group is $GL(V)=GL(9,\mathbb{C})$ and the action is still the one of \eqref{eq:sloccaction}.

The orbit structure is available due to the work of Vinberg and \'Elashvili\cite{Vinberg}. It turned out that there are a total of 164 entanglement classes. The classification is based on the unique decomposition of $P$:
\beq
P=Q+R,
\eeq
where $Q$ is semisimple and $R$ is nilpotent. States which have closed orbits under the unimodular group $SL(V)$ are called semisimple and states that have $SL(V)$ orbits whose closure contains the zero vector are called nilpotent. The 164 orbits can be grouped into seven families according to the type of their semisimple part. 

The case of nine dimensions is a bit different from the previously discussed ones. Recall that in the cases discussed so far we had only one relative invariant nonvanishing only on one particular orbit. It follows that these orbits were dense open subsets of the whole three fermions state space. We call these kinds of orbits stable. The vector spaces admitting a stable orbit when considered as a representation of a particular algebraic group are called prehomogeneous vector spaces\cite{Kimura,Satokimura}. In the case of nine dimensions we have $\dim \wedge^9 V^*=84$ and $\dim GL(9)=81$. It follows that the highest value of the local dimension of an orbit can be at most 81 thus \textit{there are no stable orbits}. It turns out\cite{Vinberg} that there are seven orbits with a zero dimensional stabilizer subgroup thus with a maximal local dimension of 81 and there are precisely one in every family. 

Now recall two standard textbook results:
\begin{enumerate}
 \item If $\phi_i$, $i=1,...,m$ are differentiable functions on a vector space $W$ then the existence of a non-trivial relation $\Omega(\phi_1,...,\phi_m)=0$ holding on an open subset $U$ of $W$ implies that the system of gradients $\lbrace \text{grad} \phi_1,...,\text{grad} \phi_m\rbrace$ is lineary dependent. Equivalently if this system is lineary independent such a relation does not exist. In this latter case we say these functions are algebraically independent on $U$. (See Proposition \ref{prop:inv1}. of Appendix \ref{app:app1}.) \
 \item If a Lie group $G$ acts on $W$ then there are at most $\dim W-\dim G-\dim \text{Stab}(v)$ algebraically independent invariant differentiable functions satisfying $\phi(v)=\phi(gv)$, $v\in W,\forall g\in G$ on $W$. Here $\text{Stab}(v)$ is the stabilizer of $v\in W$. (See Proposition \ref{prop:inv2}. of Appendix \ref{app:app1}.)
\end{enumerate}

We apply this to $W=\wedge^3 V^*$ and $G=SL(9,\mathbb{C})$. The 80 dimensional orbits obviously have $\dim \text{Stab}=0$ thus we have $84-80=4$ independent invariants w.r.t. the action of the unimodular group $SL(9,\mathbb{C})$. These are relative invariants w.r.t. the SLOCC group $GL(9,\mathbb{C})$ picking up determinant factors. Indeed, as shown previously by Vinberg\cite{Vinberg2} for this particular group and representation the algebra of invariants is freely generated by four polynomial invariants. They were first found by Egorov\cite{Egorov} with a different method from the one described here. Now we construct these invariants with the methods described in Section \ref{sec:covariants}. It turned out that this method was first used by Katanova\cite{Katanova}. Consider the covariant $\kappa_P^{(2,1)}$ with matrix elements
\beq
\label{eq:Tmatrix}
(K_P^{(2,1)})^{abc}_{\;\;\;def} \equiv T^{abc}_{\;\;\;def} = \frac{1}{2!3!}\epsilon^{abcpqrstu}P_{dep}P_{fqr}P_{stu}.
\eeq
Now because $T$ has 3 upper and 3 lower indices one can take its powers wich will have the same index structure:
\beq
\begin{aligned}
(T^2)^{a_1b_1c_1}_{\;\;\;a_3b_3c_3}  &=T^{a_1b_1c_1}_{\;\;\;a_2b_2c_2} T^{a_2b_2c_2}_{\;\;\;a_3b_3c_3} , \\
&\vdots\\
(T^m)^{a_1b_1c_1}_{\;\;\;a_{m+1}b_{m+1}c_{m+1}} &=T^{a_1b_1c_1}_{\;\;\;a_2b_2c_2} ...T^{a_mb_mc_m}_{\;\;\;a_{m+1}b_{m+1}c_{m+1}} .
\end{aligned}
\eeq
More strictly speaking by antisymmetrization of its lower indices, $T$ can be regarded as a linear map $T:\wedge^3 V\rightarrow \wedge^3 V$ hence one can compose $T$ with itself and form $T^2=T\circ T$, ...,$T^m=T\circ ...\circ T$. Now define a set of relative invariants by
\beq
\phi_{3n}=\text{Tr}T^n=(T^n)^{abc}_{\;\;\;abc}.
\eeq
The subscript $3n$ denotes the homogeneous degree of $\phi_{3n}$ in the amplitudes $P_{ijk}$. Note that $\phi_{3n}=0$ for $n$ odd and $n=2$. Let us introduce the notation
\beq
\label{eq:9egyreszinvariants}
\begin{aligned}
J_{12} &= \frac{1}{2^7 3^3 7}\phi_{12}, &&
J_{18} &= -\frac{1}{2^{10}3^3 7\cdot 13}\phi_{18},\\
J_{24} &= \frac{1}{2^{11}3^2 7\cdot 19}\phi_{24}, &&
J_{30} &= -\frac{1}{2^{12}3^3 5\cdot 7\cdot 13}\phi_{30},\\
\end{aligned}
\eeq
for later convenience.
Before proceeding with the review of the seven families and describing the properties of these invariants in each class we make an important observation.
\begin{prop}
\label{prop:nilp}
The value of any continuous $SL(V)$ invariant function is independent of the nilpotent part.
\begin{proof}
A nilpotent state by definition has the zero state in the closure of its $SL(V)$ orbit hence for $R$ nilpotent there exists a sequence $\lbrace s_n \rbrace$ in $SL(V)$ such that $s_n^*R\rightarrow 0$. Let $\phi$ be a continuous $SL(V)$ invariant and $P=Q+R$ an arbitary state. We have 
\beq
\phi(P)=\phi(Q+R)=\phi(s_n^*Q+s_n^* R).
\eeq
Taking the limit and using continuity of $\phi$ gives $\phi(P)=\phi(Q)$.
\end{proof}
\end{prop}
 Now any semisimple state $Q$ can be brougth by an $SL(V)$ transformation to the following form\cite{Vinberg}
\beq
\label{eq:semisimple}
\begin{aligned}
Q_0 &=a q_1+b q_2 +c q_3+d q_4,
\end{aligned}
\eeq
where for simplicity we use the notation
\beq
\label{eq:defofqs}
\begin{aligned}
 q_1&=E^{123}+E^{456}+E^{789}, && q_2&=E^{147}+E^{258} +E^{369}, \\
 q_3&=E^{159}+E^{267}+E^{348}, && q_4&=E^{168}+E^{249}+E^{357}.
\end{aligned}
\eeq
We give explicit expressions for the invariants $J_{12}$, $J_{18}$, $J_{24}$, $J_{30}$ evaluated at $Q_0$ in Appendix \ref{app:2}.
Define the matrix
\beq
M=
\left(
\begin{array}{cccc}
\partial_a J_{12} & \partial_a J_{18} & \partial_a J_{24} & \partial_a J_{30} \\
\partial_b J_{12} & \partial_b J_{18} & \partial_b J_{24} & \partial_b J_{30} \\
\partial_c J_{12} & \partial_c J_{18} & \partial_c J_{24} & \partial_c J_{30} \\
\partial_d J_{12} & \partial_d J_{18} & \partial_d J_{24} & \partial_d J_{30} \\
\end{array}
\right).
\eeq
Now one can check with any computer algebra system that the determinant of $M$ is a not identically zero degree 80 polynomial expression in the coefficients $a,b,c,d$:
\beq
\begin{aligned}
\frac{1}{2^{14}3^4 5^7 11^2 \cdot 6\cdot 199}\det M =& a^2 b^2 c^2 d^2\\
&((a^3 + b^3 - c^3)^3 + (3 a b c)^3)^2 \\
& ((a^3 - b^3 + d^3)^3 + (3 a b d)^3)^2 \\
&((c^3 + b^3 + d^3)^3 - (3 c b d)^3)^2 \\
&((c^3 + a^3 - d^3)^3 + (3 c a d)^3)^2.
\end{aligned}
\eeq
As a consequence, the invariants $J_{12}$, $J_{18}$, $J_{24}$, $J_{30}$ are algebraically independent on any open subset of semisimple states and hence by Proposition \ref{prop:nilp}. on any open subset of $\wedge^3 V$. We note here that the rank of the linear map $T:\wedge^3 V\rightarrow \wedge^3 V$ is also a SLOCC invariant but it is not continuous so Proposition \ref{prop:nilp}. does not apply to it. We now review the seven families of states based on where their semisimple part belong to.

\begin{itemize}
\item[]\textbf{First family}\\
This family contains only semisimple states with no possible nilpotent part. According to the work of Vinberg and \'Elashvili\cite{Vinberg} the coefficients of the canonical form \eqref{eq:semisimple} satisfy
\beq
\label{eq:familyone}
\begin{aligned}
abcd &\neq 0,\\
(b^3+c^3+d^3)^3-(3bcd)^3 &\neq 0,\\
(a^3+c^3-d^3)^3+(3acd)^3 &\neq 0,\\
(a^3-b^3+d^3)^3+(3abd)^3 &\neq 0,\\
(a^3+b^3-c^3)^3+(3abc)^3 &\neq 0.
\end{aligned}
\eeq
Notice that this is equivalent with $\det M\neq0$. Indeed, this is the only orbit with $\text{rank} M=4$. This orbit has a discrete thus zero dimensional stabilizer.  

\item[]\textbf{Second family}\\ The semisimple part has the canonical form
\beq
\label{eq:secfam}
\begin{aligned}
a q_1-b q_2 
+d q_4.
\end{aligned}
\eeq
Formaly, we can obtain this by putting $c=0$ and $b\rightarrow -b$ in \eqref{eq:semisimple}. The amplitudes satisfy\cite{Vinberg}
\beq
abd(a^3-b^3)(a^3-d^3)(b^3-d^3)((a^3+b^3+d^3)^3-(3abd)^3)\neq 0.
\eeq
Now one can check that this is equivalent to dropping the third row of $M$, putting $c=0$ and $b\rightarrow -b$ in it and requiring any of the $3\times 3$ subdeterminant of the resulting $3\times 4$ matrix to be non zero. Indeed, we have $\text{rank} M=3$ for these semisimple states. The vanishing of $\det M$ means that it is possible that a function of the four invariants exists which equals zero. It turns out that indeed there exists an invariant of degree 132 which vanishes for this family:
\beq
\begin{aligned}
\Delta_{132}&=
J_{12}^{11}-\tinyfrac{44940218765172270463 }{2232199994248855116}J_{12}^8 J_{18}^2+\tinyfrac{113325967730636958495085217 }{1009180965699898771226274}J_{12}^5 J_{18}^4\\
&-\tinyfrac{11518845901768651039 }{329340982758027804}J_{12}^2 J_{18}^6-\tinyfrac{188875 }{1526823}J_{12}^9 J_{24}+\tinyfrac{20955843759677134000 }{15067349961179772033}J_{12}^6 J_{18}^2 J_{24}\\
&-\tinyfrac{48098757899275092625 }{15067349961179772033}J_{12}^3 J_{18}^4 J_{24}+\tinyfrac{156259946875 }{27974261679948}J_{12}^7 J_{24}^2\\
&-\tinyfrac{43381098724294271875 }{2440910693711123069346}J_{12}^4 J_{18}^2 J_{24}^2-\tinyfrac{32778366465625}{48591292538069676} J_{12} J_{18}^4 J_{24}^2\\
&-\tinyfrac{37339826093750 }{327991224631970313}J_{12}^5 J_{24}^3-\tinyfrac{198339133437500}{741017211205562559} J_{12}^2 J_{18}^2 J_{24}^3\\
&+\tinyfrac{351718750000}{327991224631970313} J_{12}^3 J_{24}^4-\tinyfrac{1250000000}{327991224631970313} J_{12} J_{24}^5\\
&+\tinyfrac{522717082571600510}{5022449987059924011} J_{12}^7 J_{18} J_{30}-\tinyfrac{4631798176278228432974860}{4541314345649544470518233} J_{12}^4 J_{18}^3 J_{30}\\
&+\tinyfrac{45691574382263590}{741017211205562559} J_{12} J_{18}^5 J_{30}-\tinyfrac{951594557840795000}{135606149650617948297} J_{12}^5 J_{18} J_{24} J_{30}\\
&+\tinyfrac{2133816827644645000}{135606149650617948297} J_{12}^2 J_{18}^3 J_{24} J_{30}+\tinyfrac{140973248590625000}{1220455346855561534673} J_{12}^3 J_{18} J_{24}^2 J_{30}\\
&+\tinyfrac{10890275000000}{20007464702550189093} J_{12} J_{18} J_{24}^3 J_{30}-\tinyfrac{8007699664851700}{45202049883539316099} J_{12}^6 J_{30}^2\\
&+\tinyfrac{6686357462527147925300}{1513771448549848156839411} J_{12}^3 J_{18}^2 J_{30}^2+\tinyfrac{1392403335812500}{135606149650617948297} J_{12}^4 J_{24} J_{30}^2\\
&-\tinyfrac{2371961791512500}{135606149650617948297} J_{12} J_{18}^2 J_{24} J_{30}^2-\tinyfrac{216716472500000}{1220455346855561534673} J_{12}^2 J_{24}^2 J_{30}^2\\
&-\tinyfrac{14445540571041712000}{1513771448549848156839411} J_{12}^2 J_{18} J_{30}^3+\tinyfrac{34328756109890000}{4541314345649544470518233}J_{12} J_{30}^4.
\end{aligned}
\eeq
We have $\Delta_{132}=0$ for the second and $\Delta_{132}\neq 0$ for the first family. There are three types of possible nilpotent parts in this family. These can be found in the work of Vinberg\cite{Vinberg}. Semisimple states of this family have a two dimensional $T_2$ type stabilizer subgroup.

We note here that the second family contains general three qutrit states via the embedding of eq. \eqref{eq:quditembed}. We will discuss this in more detail in subsection \ref{sec:3qutrits}.

\item[]\textbf{Third family}\\ The canonical form of the semisimple part is
\beq
\begin{aligned}
a q_1+d q_4.
\end{aligned}
\eeq
We can obtain this by putting $b=0$ in the canonical form of the second family. The coefficients satisfy $ad(a^6-d^6)\neq 0$. There are nine types of possible semisimple parts. Semisimple states in this family have a 4 dimensional stabilizer subgroups of type $T_4$. We have $\Delta_{132}=0$ in this family.  The rank of the matrix $M$ is 2 and as one expects there exists one more function of the invariants which is identically zero in this family. Define an invariant of homogeneous degree 48 by
\beq
\begin{aligned}
\label{eq:genHDET}
\Delta_{48} &=J_{24}^2+\frac{13\cdot 23^2\cdot 293}{2^2 5^4}J_{12}^4 + \frac{3^2\cdot 11\cdot 127 \cdot 199^2}{2^3 5^4\cdot 61} J_{12}J_{18}^2\\
&-\frac{257\cdot 3^2}{5\cdot 2^3}J_{12}^2J_{24}-\frac{11\cdot 199^2}{2^2 5^3\cdot 61}J_{18}J_{30}.
\end{aligned}
\eeq
As we will explain soon, $\Delta_{48}$ is the generalization of the hyperdeterminant for $3\times 3\times 3$ arrays. We have $\Delta_{48}=0$ for states in the third family but $\Delta_{48}\neq0$ for the first and the second family.

\item[]\textbf{Fourth family}\\ The canonical form of the semisimple part is
\beq
\begin{aligned}
a q_1+b q_2- b q_3.
\end{aligned}
\eeq
Formaly, one obtains this by putting $d=0$ and $c=-b$ in \eqref{eq:semisimple}. The coefficients must satisfy $ab(a^3-b^3)(a^3+8 b^3)\neq 0$. The matrix $M$ has rank 2. We have
\beq
\Delta_{48}=2^2 \cdot 5\cdot 11^2\cdot 199^2\cdot b^9 \left(a^3-b^3\right)^9 \left(a^4+8 a b^3\right)^3
\eeq
for this family. The condition $\Delta_{48}\neq 0$ is obviously equivalent with the previous one. Of course we have $\Delta_{132}=0$ but we have another invariant of degree 48 wich vanishes here:
\beq
\label{eq:genHDET}
\begin{aligned}
\Delta_{48}' &=113\cdot 193 J_{12}^4-\frac{11\cdot 199^2\cdot 21347}{3^5 61} J_{12} J_{18}^2+\frac{2\cdot 5^3\cdot 257}{3^4} J_{12}^2 J_{24}\\
&-\frac{2^4 5^4 }{3^6}J_{24}^2+\frac{2^3\cdot 5\cdot 11 \cdot 199^2 }{3^5\cdot 61}J_{18} J_{30}.
\end{aligned}
\eeq
We have $\Delta_{48}'=0$ for the fourth family but $\Delta_{48}'\neq 0$ for the first, second and third families.
There are six types of possible nilpotent parts. The stabilizer subroup of semisimple states is of type $A_2$ and has dimension 8.

\item[]\textbf{Fifth family}\\ The canonical form of the semisimple part is
\beq
\begin{aligned}
-c q_2 +c q_3.
\end{aligned}
\eeq
This is just the canonical form of the fourth family with $a=0$. We require $c\neq 0$. The matrix $M$ has rank 1. We have $\Delta_{132}=\Delta_{48}=\Delta_{48}'=0$. There are 18 different types of possible nilpotent parts. The stabilizer subgroup is 10 dimensional and of type $A_2+T_2$ for semisimple states.

\item[]\textbf{Sixth family}\\ The canonical form of the semisimple part is
\beq
\begin{aligned}
a q_1,
\end{aligned}
\eeq
with $a\neq 0$. This is just the state $\eqref{eq:semisimple}$ with $b=c=d=0$. The matrix $M$ has rank 1 and $\Delta_{132}=\Delta_{48}=\Delta_{48}'=0$. Moreover it is easy to see that the degree 24 invariant
\beq
\Delta_{24}=J_{12}^2-\frac{1}{111}J_{24}
\eeq
is zero for this family while it is non-zero for families 1-5.
There are 25 different types of possible nilpotent parts. The semisimple states in this family have a 24 dimensional stabilizer subgroup of type $3A_2$.

\item[]\textbf{Seventh family}\\ The semisimple part is zero here thus this family is the family of nilpotent states. By Proposition \ref{prop:nilp}. all the continuous invariants are zero here. There are 102 different types of nilpotent states listed in the work of Vinberg and \'Elashvili\cite{Vinberg}. When considered as states of the nine dimensional system, all states with a lower dimensional single particle Hilbert space discussed in the previous sections are in this family.
\end{itemize}

\begin{table}[h!]
\centering
\begin{tabular}{|c|c|c|c|c|c|}
\hline
Family & $\Delta_{132}$ & $\Delta_{48}$ & $\Delta_{48}'$ & $\Delta_{24}$ & Rank $T$\\
 \hline \hline
First & $\neq 0$ & $\neq 0$ & $\neq 0$ & $\neq 0$ & 80\\
Second & 0 & $\neq 0$ & $\neq 0$ & $\neq 0$ &78\\
Third & 0 & 0 & $\neq 0$ & $\neq 0$ & 76\\
Fourth & 0 & $\neq 0$ & 0 & $\neq 0$ & 72\\
Fifth & 0 & 0 & 0 & $\neq 0$ & 70\\ 
Sixth & 0 & 0 & 0 & 0 & 56\\
\hline
\end{tabular}
\caption{Values of the new continuous invariants on families of three fermions with nine single particle states. The last column is the rank of the map $T=\kappa^{(2,1)}$ on semisimple states with zero nilpotent part.}
\label{tab:4}
\end{table}
A summary of the families and their resolution with the invariants $\Delta_{132},\Delta_{48},\Delta_{48}',\Delta_{24}$ can be found in TABLE \ref{tab:4}. 

Recall that every previously discussed case had a stable, ``GHZ-like'' orbit with non zero value of an invariant. Intuitively speaking, we have more than one ``GHZ-like'' orbits in the present case. These are families 1-6 which have at least one invariant with a non zero value. The nilpotent orbits of the seventh family can be thought of as non-GHZ like orbits where all invariants vanish. In the first family there is no possible way of combining zero out of the four invariants. By this property one can think of the first family as the ``most GHZ-like'' orbit of the ``GHZ-like'' orbits.

Before moving on to the discussion of the embedded three qutrit system we would like to make an interesting observation. The rank of the linear map $T:\wedge^3V \rightarrow \wedge^3 V$ defined in \eqref{eq:Tmatrix} is just
\beq
\text{rank} T=80-\dim \text{stab}(Q)
\eeq
for semisimple states $Q$ with zero nilpotent part.

\subsubsection{Entanglement of three qutrits}
\label{sec:3qutrits}

A qutrit is a three state quantum system with Hilbert space $\mathcal{H}=\mathbb{C}^3$. The Hilbert space of three distinguishable qutrits is just $\mathcal{H}^{\otimes 3}\cong \mathbb{C}^9$. With $\lbrace |1\rangle, |2\rangle, |3\rangle \rbrace$ being a basis of $\mathcal{H}$ a general 3 qutrit state can be written as
\beq
|\psi\rangle=\sum_{\mu_1,\mu_2\mu_3=1}^3\psi_{\mu_1\mu_2\mu_3}|\mu_1\mu_2\mu_3\rangle.
\eeq
The SLOCC group is $GL(3,\mathbb{C})^{\times 3}$ and it acts on the 9 complex amplitudes as
\beq
\psi_{\mu_1\mu_2\mu_3} \mapsto (S_1)_{\mu_1}^{\;{\nu_1}}(S_2)_{\mu_2}^{\;{\nu_2}}(S_3)_{\mu_3}^{\;{\nu_3}}\psi_{\nu_1\nu_2\nu_3}, \;\; S_1\otimes S_2\otimes S_3 \in GL(3,\mathbb{C})^{\times 3}.
\eeq
The mathematical problem of finding the SLOCC classes was solved by Nurmiev\cite{Nurmiev1,Nurmiev2}. Explicit expressions for the three continuous invariants generating the invariant algebra of this system was found by Briand et al.\cite{Briand} where the problem was also recognized as the problem of SLOCC classification of three qutrits. Later Bremner and Hu managed to express the hyperdeterminant\cite{Gelfand} of a $3\times 3\times 3$ array with these three invariants\cite{Bremner,Bremner2}. In the following we identify the problem of three qutrit entanglement as a special case of entanglement of three fermions with nine single particle states. We relate the invariants $I_6,I_9,I_{12}$ and the hyperdeterminant $\Delta_{333}$ of Bremner and Hu with the invariants of eq. \eqref{eq:9egyreszinvariants}.

According to Nurmiev\cite{Nurmiev1,Nurmiev2} any $3\times 3\times 3$ array can be uniquely written as the sum of a semisimple and a nilpotent part. Just like in the case of three fermions a semisimple state is defined to have a closed $SL(3,\mathbb{C})^{\times 3}$ orbit while a nilpotent state has the zero vector in the closure of its orbit. Now any semisimple state can be brought to a so called normal form:
\beq
|\psi_0\rangle = a|X_1\rangle-b|X_2\rangle +c|X_3\rangle,
\eeq
where
\beq
\begin{aligned}
|X_1\rangle &=|111\rangle +|222\rangle + |333\rangle, && |X_2\rangle&=|123\rangle +|231\rangle +|312\rangle, \\
|X_3\rangle &= |132\rangle +|213\rangle +|321\rangle.
\end{aligned}
\eeq
There are a total of 43 orbits under the action of $GL(3,\mathbb{C})^{\times 3}$ and these can be grouped into five families according to the type of their semisimple part. The three fundamental invariants evaluated at the normal form $|\psi_0\rangle$ are\cite{Bremner2}:
\beq
\begin{aligned}
I_6&= a^6+10 a^3 b^3+b^6-10 a^3 c^3+10 b^3 c^3+c^6, \\
I_9 &=(a+b) (a-c) (b+c)\left(a^2-a b+b^2\right)  \left(a^2+a c+c^2\right) \left(b^2-b c+c^2\right),\\
I_{12} &=-a^9 b^3-4 a^6 b^6-a^3 b^9+a^9 c^3-2 a^6 b^3 c^3+2 a^3 b^6 c^3-b^9 c^3-4 a^6 c^6\\
&-2 a^3 b^3 c^6-4 b^6 c^6+a^3 c^9-b^3 c^9.
\end{aligned}
\eeq
The hyperdeterminant for $3\times 3\times 3$ arrays has degree 36 and expressed with these invariants as\cite{Bremner2}:
\beq
\Delta_{333}=I_{6}^3 I_{9}^2-I_{12}^2 I_{6}^2-32 I_{12}^3+36 I_{12} I_{6} I_{9}^2+108 I_{9}^4.
\eeq
It has the propierty that it is zero for all families except the first one. Now consider the map defined in \eqref{eq:quditembed} for $d=k=3$ and denote it by $\chi: \mathbb{C}^3\otimes \mathbb{C}^3 \otimes \mathbb{C}^3 \rightarrow \wedge^3 (\mathbb{C}^9)^*$:
\beq
\label{eq:chi}
\begin{aligned}
\chi:& |\psi\rangle \mapsto
P_\psi &= \sum_{\mu_1,\mu_2,\mu_3=1}^3 \psi_{\mu_1\mu_2\,\mu_3} e^{\mu_1}\wedge e^{3+\mu_2} \wedge e^{6+\mu_3}. 
\end{aligned}
\eeq
Now it is very easy to check that 
\beq
\begin{aligned}
\chi(|X_1\rangle)&=P_{X_1} = q_2, && \chi(|X_2\rangle)&=P_{X_2} = q_3, && 
\chi(|X_3\rangle)&=P_{X_3} = q_4,
\end{aligned}
\eeq
where $q_1,...,q_4$ are defined in e.q. \eqref{eq:defofqs}.

\begin{figure}[h!]
\includegraphics[width=0.9\textwidth]{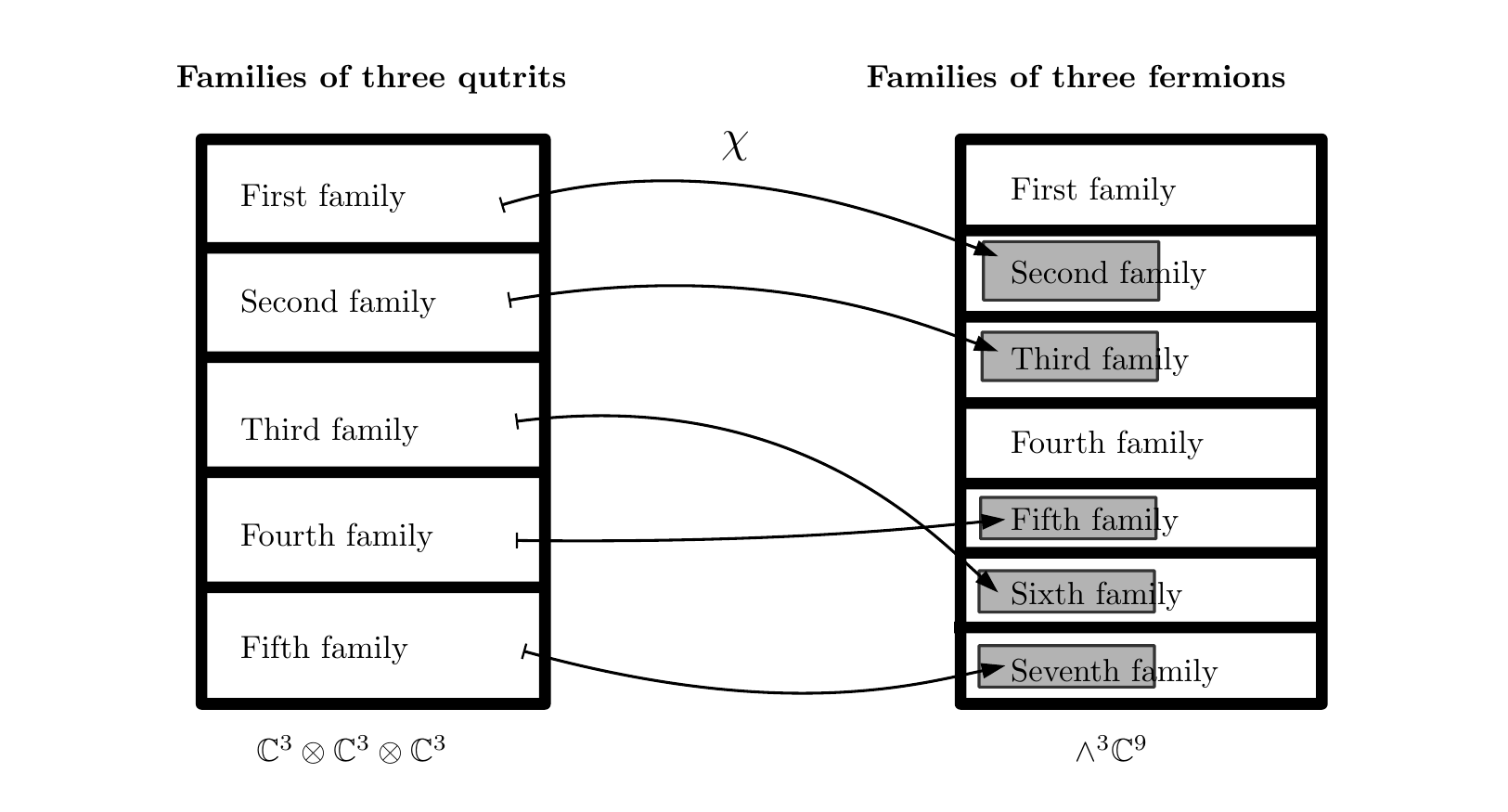}
\caption{A sketch showing how the embedding $\chi:\mathbb{C}^3\otimes \mathbb{C}^3 \otimes \mathbb{C}^3 \rightarrow \wedge^3 (\mathbb{C}^9)^*$ defined in \eqref{eq:chi} works. The grey rectangles on the right side represent the image of the three qutrit families under $\chi$. Different families are mapped into different families.}
\label{fig:3}
\end{figure}

\begin{prop}
\label{prop:invariants}
 On $\text{Im}\chi \subset \wedge^3 V$ the invariants of \eqref{eq:9egyreszinvariants} can be expressed with the fundamental invariants of three qutrits as
\beq
\begin{aligned}
J_{12}&=I_6^2+20 I_{12}, \\
J_{18}&=I_6^3+30 I_{12} I_6 + 100 I_9^2, \\
J_{24}&=111 I_6^4+4440 I_6^2 I_{12} +2\cdot 3^4\cdot 193 I_{12}^2 + 2^2\cdot  11\cdot 199 I_6 I_9^2,\\
J_{30}&=2\cdot 3^2\cdot 5^2 \cdot 2521 I_9^2 I_{12} + 3^3\cdot 5\cdot 2521 I_6 I_{12}^2 \\
&+ 2\cdot 5\cdot 17\cdot 383 I_6^2 I_9^2 + 2^4\cdot 5^2\cdot 73 I_6^3 I_{12} + 
 2^3\cdot 73 I_6^5.
\end{aligned}
\eeq
Moreover the invariant $\Delta_{48}$ is expressed with the hyperdeterminant as
\beq
\Delta_{48}=-\frac{5\cdot 11^2\cdot 199^2}{2}\Delta_{333} I_{12}.
\eeq
For the other $\Delta$ invariants we have:
\beq
\begin{aligned}
\Delta_{138} &=0, \\
\Delta_{48}'&=\frac{2^4 5^5  11^2 199^2}{3^5 }\left(2^3 I_{12} + \frac{1}{3} I_6^2 \right)I_9^4, \\
\Delta_{24}&=\frac{2\cdot 11\cdot 199}{37}\left(I_{12}^2-\frac{2}{3} I_6 I_9^2\right).
\end{aligned}
\eeq
\begin{proof}
The relations can be checked with any computer algebra system for the image of the normal form $P_{\psi_0}$ which is actually the canonical form \eqref{eq:secfam} of the second family. By invariance they are true for any semisimple state. By Proposition \ref{prop:nilp}. they remain true by adding any nilpotent state.
\end{proof}
\end{prop}
Define
\beq
\begin{aligned}
 D_{36}&=\Delta_{333}, && D_{24}&=I_{12}^2-\frac{2}{3} I_6 I_9^2, &&
 D_{21}&= (2^3 I_{12} + \frac{1}{3} I_6^2)I_9.
\end{aligned}
\eeq
As a consequence of Proposition \ref{prop:invariants}. the invariants $D_{36},D_{24},D_{21}$ completely separate the five families of three qutrits. One can find representatives of these five families e.g. in the work of Bremner et. al.\cite{Bremner2}. We followed the enumeration of the families used there. The first family has $D_{36}\neq 0$, the second family has $D_{36}=0,D_{24}\neq0, D_{21}\neq 0$, the third family has $D_{36}=D_{24}=D_{21}=0$ and finally the fourth family has $D_{36}=D_{21}=0, D_{24}\neq 0$. For the nilpotent orbits of the fifth family every fundamental invariant vanishes. On FIG. \ref{fig:3}. we sketched how the embedding $\chi$ works. The images of different families are disjoint.

\section{Pinning of occupation numbers}
\label{sec:pinning}

As a possibly relevant physical application we would like to comment on a connection of the above SLOCC classification of fermionic quantum states with the Klyachko constraints\cite{Kly1} on the eigenvalues of the one particle reduced density matrix (or one-matrix). These constraints define a polytope in the space of possible eigenvalues of the one-matrix. An important concept is the so called pinning of occupation numbers which is the saturation of these Klyachko constraints\cite{Kly2,Christandl2}. It is widely believed that energy minima of many fermion systems usually do not lie in the Klyachko polytope thus the ground state will be on the boundary of the polytope hence it will be pinned. Indeed there are both analytical\cite{Christandl2} and numerical\cite{Benavides} results that such a pinning occurs in ground states of realistic systems. As shown by Klyachko\cite{Kly2} pinning of a state imposes selection rules on it reducing the number of separable states or Slater 
determinants that it contains. This is particulary 
usefull in molecular physics since it simplifies the form of the ansatz one must use in variational methods to find 
the ground state. 

Consider first the case of three fermions with six single particle states discussed in Section \ref{sec:threefsixs}. The classical Borland-Dennis result\cite{Borland} is that if one orders the eigenvalues of the one-matrix as $\lambda_{i+1}\geq \lambda_i$ then one has a non-trivial inequality
\beq
\label{eq:Borland}
\lambda_5+\lambda_6\geq \lambda_4.
\eeq
Note that this inequality is independent of the normalization of the original pure state.
Now if \eqref{eq:Borland} is saturated for a state $P$ then it must have the form\cite{Christandl2,Kly2}
\beq
\label{eq:pinnablestate6}
P=\alpha e^1\wedge e^2\wedge e^3 +\beta e^1\wedge e^4\wedge e^5 + \gamma e^2\wedge e^4\wedge e^6,
\eeq
in the basis of natural orbitals. Natural orbitals are the eigenvectors of the one particle reduced density matrix $\rho_P$ thus we have $\rho_P e^i=\lambda_i e^i$. It is clear that transforming an arbitary state to its natural orbital form amounts to a local unitary transformation hence it does not change the SLOCC class of it. Now if we calculate our covariant $K_P^{[1,1]}$ for the state \eqref{eq:pinnablestate6} we get
\beq
\left(
\begin{array}{cccccc}
 0 & 0 & 0 & 0 & 0 & 0 \\
 0 & 0 & 0 & 0 & 0 & 0 \\
 0 & 0 & 0 & -2\beta  \gamma  & 0 & 0 \\
 0 & 0 & 0 & 0 & 0 & 0 \\
 0 & 2\alpha  \gamma & 0 & 0 & 0 & 0 \\
 -2\alpha  \beta & 0 & 0 & 0 & 0 & 0
\end{array}
\right).
\eeq
This matrix has rank 3, 1 or 0 depending on the value of the coefficients. Looking at TABLE \ref{tab:1}. we already conclude that \textit{pinning is impossible for states in the GHZ class} or otherwise stated pinning is impossible for states with $\mathcal{D}(P)=\frac{1}{6} \text{Tr}( K_P^{[1,1]})^2\neq 0$ (see eq. \eqref{Threetanglegen}). One might think that this means that all states with $\mathcal{D}(P)=0$ are pinned but this is not the case since the spectrum of the one-matrix is not invariant under general SLOCC transformations thus pinning is not a SLOCC invariant concept. Indeed one can easily find both pinned and unpinned states in the W class. Note that these observations are in perfect agreement with the numerical work done by C. L. Benavides-Riveros et. al.\cite{Benavides} where pinning was studied in finite rank variational approximations of the ground state of lithium. It was observed there that pinning in the rank six approximation can only occur if the invariant $\mathcal{D}(P)$ is zero.

Consider now the case of seven single particle states of Section \ref{sec:threefseven}. Pinning for this system is investigated by Klyachko as it is important in studying the first excited state of beryllium\cite{Kly2}. Moreover it is used as the rank 7 approximation of lithium orbitals where pinning was also observed\cite{Benavides}. For seven single particle states we have four non-trivial Klyachko constraints:
\beq
\label{eq:7constr}
\begin{aligned}
\lambda_1+\lambda_2+\lambda_4+\lambda_7 \leq 2, \\
\lambda_1+\lambda_2+\lambda_5+\lambda_6 \leq 2, \\
\lambda_2+\lambda_3+\lambda_4+\lambda_5 \leq 2, \\
\lambda_1+\lambda_3+\lambda_4+\lambda_6 \leq 2. \\
\end{aligned}
\eeq
Suppose we saturate the first one: $\lambda_1+\lambda_2+\lambda_4+\lambda_7=2$ for a normalized state $\mathcal{P}$. Then the arising selection rules imply\cite{Kly2} that $\mathcal{P}\in \wedge^2 \mathbb{C}^4 \otimes \mathbb{C}^3 \subset \wedge^3 \mathbb{C}^7$. In particular in the basis of natural orbitals $\mathcal{P}$ must be a linear combination of separable states with two indices from the set $\lbrace 1,2,4,7\rbrace$ and one index from the set $\lbrace 3,5,6\rbrace$. One can easily calculate the covariants $N^{AB}$ and $(M^{AB})_C$ of eqs. \eqref{enformula} and \eqref{emformula} for such states  and conclude that  Rank $\kappa^{(1,1)}_\mathcal{P}=4$ and Rank $\kappa^{(1)}_\mathcal{P}=7$. Looking at TABLE \ref{tab:2}. we already deduce that there is no pinning for states in class X or equivalently for states with a non-vanishing $\mathcal{J}(\mathcal{P})$ invariant (see eq. \eqref{sevenrelinv}). 

Now suppose we saturate three (the first two and the last one in e.q. \eqref{eq:7constr}) of the constraints. In this case $\mathcal{P}$ must have the form\cite{Kly2}
\beq
\label{eq:7statetotalpinned}
\mathcal{P}=\alpha e^1\wedge e^2\wedge e^3 +\beta e^1\wedge e^4\wedge e^5 +\gamma e^1\wedge e^6\wedge e^7+  \delta e^2\wedge e^4\wedge e^6,
\eeq
when expanded on its natural orbitals. Calculating the relevant ranks for this state gives Rank $\kappa^{(1,1)}_\mathcal{P}=1$ and Rank $\kappa^{(1)}_\mathcal{P}=4$ which identifies class VII of TABLE \ref{tab:2}. Moreover one can check that one cannot increase the rank of $\kappa^{(1)}_\mathcal{P}$ by setting any of the coefficients to zero. This means that states of the form \eqref{eq:7statetotalpinned} cannot be in class V. However, they do cross classes VII,VI,IV,III,II and I of TABLE \ref{tab:2}. so we deduce that pinning of three Klyachko constraints is only possible for states in these classes and impossible in classes V,VIII,IX and X. If we require the saturation of all four constraints then we have to put $\gamma =0$ in \eqref{eq:7statetotalpinned} and we get back to a state of the form \eqref{eq:pinnablestate6}. Thus pinning of all four constraints is only possible in a six single particle subspace and only in the classes I-IV. 

\section{Conclusions}
\label{sec:conclusions}

In this work we presented a method 
to generate SLOCC covariants and invariants for multifermion systems. Based on results taken from the mathematical literature we have presented the SLOCC classification of three fermions with 6, 7, 8 and 9 single particle states.
We also discussed how this classification can be understood with the help of covariants and invariants. In the special cases of six and seven dimensions we managed to characterize the SLOCC entanglement classes geometrically via mapping the canonical forms of the classes to special subconfigurations of the Fano plane. We have also revealed that in the $6,7,8$ dimensional cases the classes giving rise to stable orbits are examples of prehomogeneous vector spaces\cite{Kimura,Satokimura}. For these classes there is a characteristic relative invariant which is nonvanishing.  In all of these cases these classes are giving rise to dense, Zariski-open orbits with representatives playing a role similar to the classical three qubit GHZ state. In the cases of six and seven single particle states we outlined some connections between the discussed SLOCC classification and  the celebrated Klyachko constraints on the spectra of one particle reduced dentsity matrices. In particular we observed that saturation (or pinning) 
of the constraints is not possible in every SLOCC class.

In the case of nine dimensions there is no stable orbit and there are four algebraically independent polynomial invariants. The SLOCC 
orbits can be organized into seven families. The seventh family contains nilpotent orbits where all of the four invariants vanish. These can be considered as a generalization of the non-GHZ classes. The rest of the families have at least one invariant with a non zero value thus these can be thought of as a generalization of GHZ-like orbits. We have shown that these families can be distinguished via a calculation of an order 132, two order 48 and an order 24 combination of the fundamental invariants. We have also shown that the entanglement classification of three qutrits and the corresponding invariant algebra can be recovered from the embedding of the system into the one of three fermions with nine single particle states. In particular the $3\times 3\times 3$ hyperdeterminant arises as a factorization of one of the invariants of order 48.

\appendix

\section{Proof of some theorems}
\label{app:app1}

Here we present some textbook theorems for Section \ref{sec:threefnine}. and some explicit calculations leading to some of the results of Section \ref{sec:threefseven}.

\begin{prop}
\label{prop:inv1}
Let $M$ be an $n$ dimensional manifold. There are at most $n$ algebraicly independent functions on $V$.
\begin{proof}
Let $\phi_i :M\rightarrow \mathbb{C}$, $i=1,...,m$ functions on $M$. Suppose that there exists $\Omega : \mathbb{C}^m \rightarrow \mathbb{C}$ such that
\beq
\Omega(\phi_1(x),...,\phi_m(x))=0,
\eeq
on every point $x\in U$ of an open subset $U\subset M$. Taking the exterior derivative with coordinates $x^a$, $a=1,...,n$ yields
\beq
d\Omega=\left( \frac{\partial \Omega}{\partial\phi_1}\frac{\partial \phi_1(x)}{\partial x^a}+...+\frac{\partial \Omega}{\partial\phi_m}\frac{\partial \phi_m(x)}{\partial x^a}\right) dx^a=0,
\eeq
hence if $\Omega$ has nonvanishing derivatives at $\phi_i(x)$ then the system of $n$ component vectors $\lbrace \frac{\partial \phi_1}{\partial x^a},...,\frac{\partial \phi_m}{\partial x^a}\rbrace$ is lineary dependent. The negation of the above result reads as: if the system $\lbrace \frac{\partial \phi_1}{\partial x^a},...,\frac{\partial \phi_m}{\partial x^a}\rbrace$ is lineary independent on $U$, then $\Omega(\phi_1,...,\phi_m)=0$ possible only if $\Omega$ is constant zero on $U$.
\end{proof}
\end{prop}

\begin{prop}
\label{prop:inv2}
Let $M$ be an $n$ dimensional manifold, $G$ a Lie group with a group action $\rho$ on $M$ and $\phi: M\rightarrow \mathbb{C}$ a differentiable $G$-invariant function i.e. $\phi(x)=\phi(\rho(g)x)$, $\forall x\in M,\forall g\in G$. There are at most $n-\dim G + \dim H_x$ algebraicly independent such functions at a point $x\in M$, where $H_x$ is the stabilizer of $x$.
\begin{proof}
Take $g=\text{exp}(\epsilon t)$, $t\in \mathfrak{g}$ in $\phi(x)=\phi(\rho(g)x)$. Then take the derivative w.r.t $\epsilon$ and put $\epsilon=0$ to obtain
\beq
d\phi(x)(V_t)=\frac{\partial \phi(x)}{\partial x^a}(V_t)^a =0,
\eeq
where $(V_t)^a$ are the components of the tangent vector $V_t=\frac{d}{d\epsilon}\rho(\text{exp}(\epsilon t))x|_{\epsilon=0}\in T_xM$. Now since $H_x$ is a subgroup its Lie algebra $\mathfrak{h}_x$ is a linear subspace in $\mathfrak{g}$ hence $\mathfrak{g}=\mathfrak{h}_x\oplus \mathfrak{m}_x$ as vector spaces. If $t\in \mathfrak{h}_x$ then the above is automatically satisfied but if $t\in \mathfrak{m}_x$ then $(V_t)^a\neq 0$ and the above means that $\frac{\partial \phi(x)}{\partial x^a}$ is in the orthogonal complement of the space $M_x=\text{Span}\lbrace V_t|t\in \mathfrak{m}_x \rbrace\subset T_x M$. It is easy to see that $\dim M_x = \dim \mathfrak{m}_x=\dim G -\dim H_x$. Taken together with Proposition \ref{prop:inv1}. the claim follows.
\end{proof}

\end{prop}

Some detailed calculations for Section \ref{sec:threefseven}.:

\begin{prop}
\label{prop:D1} Let $\tilde{P}=\frac{1}{3!}{K^c}_iP_{cjk}e^i\wedge
e^j\wedge e^k.$ Then $P\wedge \omega=0$ implies $\tilde{P}\wedge
\omega=0$.
\begin{proof}
We know that ${\rm Tr}(K)=0$ hence $K$ is an element of the Lie
algebra of $SL(6,\mathbb{C})$. The action of $K$ on the five form
$P\wedge \omega$ is\cite{Gualtieri}

\beq K\cdot (P\wedge\omega)=-\frac{1}{2}{\rm Tr}(K)P\wedge
\omega+{K^a}_be^b\wedge
\iota_{e_a}(P\wedge\omega)={K^a}_be^b\wedge
\iota_{e_a}P\wedge\omega-{K^a}_be^b\wedge
P\wedge\iota_{e_a}\omega. \eeq Now \beq
\iota_{e_a}P\wedge\omega=\frac{1}{4}P_{aij}\omega_{kl}e^i\wedge
e^j\wedge e^k\wedge e^k,\qquad
\iota_{e_a}\omega=\omega_{al}e^l\eeq hence \beq 0=K\cdot(P\wedge
\omega)=\left(\frac{1}{4}{K^a}_bP_{aij}\omega_{kl}-({K^a}_b\omega_{al})P_{ijk}\right)
e^b\wedge e^i\wedge e^j\wedge e^k\wedge e^l.\eeq According to
Eq.(\ref{nesek2v})
${K^a}_b\omega_{al}=\frac{1}{3}N_{bl}=\frac{1}{3}N_{lb}$. Since
$N_{bl}$ is symmetric and $e^b\wedge e^l$ is antisymmetric the
second term gives zero. Hence the first term which is proportional
to $\tilde{P}\wedge \omega$ vanishes as claimed.
\end{proof}
\end{prop}

\begin{prop}
\label{prop:D2} If $P\wedge\omega =0$ then \beq
\left(P_{abi}\omega_{jk}+\frac{1}{3}\omega_{ab}P_{ijk}+P_{aij}\omega_{bk}-P_{bij}\omega_{ak}\right)e^{ijk}=0
\label{ezthasznaljuk} \eeq
\begin{proof}
The result immediately follows from the identity \beq
\iota_{e_a}\iota_{e_b}(P\wedge\omega)=\iota_{e_a}\iota_{e_b}P\wedge\omega
+\iota_{e_b}P\wedge\iota_{e_a}\omega-\iota_{e_a}P\wedge\iota_{e_b}\omega+P\wedge\iota_{e_a}\iota_{e_b}\omega=0.
\eeq
\end{proof}
\end{prop}

\begin{prop}
\label{prop:D3} Let $L^{AB}\equiv {(M^A)^C}_D{(M^B)^D}_C$ and
$P\wedge \omega=0$. Then we have $L^{77}=6\mathcal{D}(P)$ and
$L^{7a}=L^{a7}=0$.
\begin{proof} By virtue of Eq. (\ref{mesek1v}) and (\ref{Threetanglegen}) we have
\beq L^{77}={(M^7)^C}_D{(M^7)^D}_C= {(M^7)^c}_d{(M^7)^d}_c=
{K^c}_d{K^d}_c=6\mathcal{D}(P).\eeq On the other hand using
Eq.(\ref{mesek2v}) one gets \beq
L^{7a}=L^{a7}={(M^7)^c}_d{(M^a)^d}_c={K^c}_d\frac{1}{2}\varepsilon^{adijkl}P_{cij}\omega_{kl}=
\frac{1}{2}\varepsilon^{adijkl}\tilde{P}_{dij}\omega_{kl}. \eeq
Using Proposition \ref{prop:D1}. the latter expression is zero.
\end{proof}
\end{prop}

\begin{prop}
\label{prop:D4} If $P\wedge\omega =0$ then
$L^{ab}=\frac{3}{2}{K^a}_c\varepsilon^{cbijkl}\omega_{ij}\omega_{kl}
=\frac{3}{2}{K^b}_c\varepsilon^{caijkl}\omega_{ij}\omega_{kl}$
\begin{proof}
\beq
L^{ab}={(M^a)^7}_d{(M^b)^d}_7+{(M^a)^c}_7{(M^b)^7}_c+{(M^a)^c}_d{(M^b)^d}_c
\label{kellezisittni} \eeq
 \beq
 {(M^a)^7}_d{(M^b)^d}_7=\frac{1}{4}{K^a}_d\varepsilon^{dbijkl}\omega_{ij}\omega_{kl},\qquad
{(M^a)^c}_7{(M^b)^7}_c=\frac{1}{4}{K^b}_d\varepsilon^{daijkl}\omega_{ij}\omega_{kl}.
 \eeq
 \beq
 {(M^a)^c}_d{(M^b)^d}_c=\frac{1}{4}\varepsilon^{acijkl}\varepsilon^{bdmnrs}P_{dij}P_{cmn}\omega_{kl}\omega_{rs}.
 \eeq
Now using Proposition \ref{prop:D2}. in the last term one can
write \beq
\varepsilon^{bmndrs}P_{ijd}\omega_{rs}=\left(-\frac{1}{3}\omega_{ij}P_{drs}+P_{idr}\omega_{js}-P_{jdr}\omega_{is}\right)
\varepsilon^{bmndrs}.\eeq Using this one can write \beq
{(M^a)^c}_d{(M^b)^d}_c={K^b}_d\varepsilon^{daijkl}\omega_{ij}\omega_{kl}+\frac{1}{2}\varepsilon^{acijkl}
\varepsilon^{bmndrs}\omega_{kl}\omega_{js}P_{cmn}P_{idr}. \eeq Now
since in the first Levi-Civita symbol we have antisymmetry in the
indices $(c,i)$ and in the second Levi-Civita symbol we have
symmetry in the pair of indices $(mn,dr)$ the last term is zero.
Using the symmetry of $G^{ab}$ the three different terms of
Eq.(\ref{kellezisittni}) gives the same type of terms with a
prefactor of $3/2$. Notice that using the definition of
$\tilde{\omega}$ of Eq.(\ref{omegatilde}) one can write \beq
L^{ab}=-12\tilde{\omega}^{ac}{K^b}_c=-12\tilde{\omega}^{bc}{K^a}_c.\label{elabe}
\eeq \noindent This result taken together with the ones of
Proposition \ref{prop:D3}. yields the factorized form for
$\boldsymbol{L}$ of Eq.(\ref{elfaktorosan}).
\end{proof}
\end{prop}

\section{Explicit expressions for the four independent invariants of three fermions with nine single particle states}
\label{app:2}

Here we list explicit expressions for the invariants of eq. \eqref{eq:9egyreszinvariants} of Section \ref{sec:threefnine}. evaluated on the canonical form \eqref{eq:semisimple} of semisimple states.
\beq
\begin{aligned}
 J_{12} &= a^{12}+b^{12}+c^{12}+22 c^6 d^6+d^{12}-220 a^3 (b^3-c^3) (b^3-d^3) (c^3-d^3) \\
&+ 220 b^3 c^3 d^3 (c^3+d^3) +22 b^6 (c^6+10 c^3 d^3+d^6)\\
& +22 a^6 (b^6+c^6-10 c^3 d^3+d^6-10 b^3 (c^3+d^3)),
\end{aligned}
\eeq
\beq
\begin{aligned}
 J_{18} &= a^{18}+b^{18}+c^{18}-17 c^{12} d^6-17 c^6 d^{12}+d^{18}+1870 a^9 (b^3-c^3) (b^3-d^3) (c^3-d^3)\\ 
 &-1870 b^9 c^3 d^3 (c^3+d^3)-17 b^{12} (c^6+10 c^3 d^3+d^6)\\
 &-170 b^3 c^3 d^3 (c^9+11 c^6 d^3+11 c^3 d^6+d^9)-17 b^6 (c^{12}+110 c^9 d^3+462 c^6 d^6+110 c^3 d^9 +d^{12})\\
 &-17 a^{12} (b^6+c^6-10 c^3 d^3+d^6-10 b^3 (c^3+d^3))-17 a^6 (b^{12}+c^{12}\\
 &-110 c^9 d^3+462 c^6 d^6-110 c^3 d^9+d^{12}-110 b^9 (c^3+d^3)+462 b^6 (c^6
 \\ &+d^6)-110 b^3 (c^9+d^9))+170 a^3 (b^{12} (c^3-d^3)-11 b^9 (c^6-d^6)\\
 &+11 b^6 (c^9-d^9)+c^3 d^3 (c^9-11 c^6 d^3+11 c^3 d^6-d^9)+b^3 (-c^{12}+d^{12})),
\end{aligned}
\eeq
\beq
\begin{aligned}
J_{24} &= 111 a^{24}+111 b^{24}+111 c^{24}+506 c^{18} d^6+10166 c^{12} d^{12}+506 c^6 d^{18}\\
&+111 d^{24}-206448 a^{15} (b^3-c^3) (b^3-d^3) (c^3-d^3)+206448 b^{15} c^3 d^3 (c^3+d^3)\\
&+506 b^{18} (c^6+10 c^3 d^3+d^6)+1118260 b^9 c^3 d^3 (c^9+11 c^6 d^3+11 c^3 d^6+d^9)\\
&+10166 b^{12} (c^{12}+110 c^9 d^3+462 c^6 d^6+110 c^3 d^9+d^{12})+1012 b^3 c^3 d^3 (5 c^{15}\\
&+204 c^{12} d^3+1105 c^9 d^6+1105 c^6 d^9+204 c^3 d^{12}+5 d^{15})+506 b^6 (c^{18}\\
&+408 c^{15} d^3+9282 c^{12} d^6+24310 c^9 d^9+9282 c^6 d^{12}+408 c^3 d^{15}+d^{18})\\
&+506 a^{18} (b^6+c^6-10 c^3 d^3+d^6-10 b^3 (c^3+d^3))+10166 a^{12} (b^{12}+c^{12}-110 c^9 d^3\\
&+462 c^6 d^6-110 c^3 d^9+d^{12}-110 b^9 (c^3+d^3)+462 b^6 (c^6+d^6)-110 b^3 (c^9+d^9))\\
&-1118260 a^9 (b^{12} (c^3-d^3)-11 b^9 (c^6-d^6)+11 b^6 (c^9-d^9)+c^3 d^3 (c^9-11 c^6 d^3+11 c^3 d^6\\
&-d^9)+b^3 (-c^{12}+d^{12}))+506 a^6 (b^{18}+c^{18}-408 c^{15} d^3+9282 c^{12} d^6-24310 c^9 d^9\\
&+9282 c^6 d^{12}-408 c^3 d^{15}+d^{18}-408 b^{15} (c^3+d^3)+9282 b^{12} (c^6+d^6)\\
&-24310 b^9 (c^9+d^9)+9282 b^6 (c^{12}+d^{12})-408 b^3 (c^{15}+d^{15}))\\
&-1012 a^3 (5 b^{18} (c^3-d^3)-204 b^{15} (c^6-d^6)+1105 b^{12} (c^9-d^9)-1105 b^9 (c^{12}-d^{12})\\
&+c^3 d^3 (5 c^{15}-204 c^{12} d^3+1105 c^9 d^6-1105 c^6 d^9+204 c^3 d^{12}-5 d^{15})\\
&+204 b^6 (c^{15}-d^{15})-5 b^3 (c^{18}-d^{18})) ,
\end{aligned}
\eeq
\beq
\begin{aligned}
 J_{30}&=  584 a^{30}+584 b^{30}+584 c^{30}-435 c^{24} d^6-63365 c^{18} d^{12}-63365 c^{12} d^{18}\\
 &-435 c^6 d^{24}+584 d^{30}+440220 a^{21} (b^3-c^3) (b^3-d^3) (c^3-d^3)\\
 &-440220 b^{21} c^3 d^3 (c^3+d^3)-435 b^{24} (c^6+10 c^3 d^3+d^6)\\
 &-25852920 b^{15} c^3 d^3 (c^9+11 c^6 d^3+11 c^3 d^6+d^9)-63365 b^{18} (c^{12}+110 c^9 d^3\\
 &+462 c^6 d^6+110 c^3 d^9+d^{12})-1394030 b^9 c^3 d^3 (5 c^{15}+204 c^{12} d^3+1105 c^9 d^6\\
 &+1105 c^6 d^9+204 c^3 d^{12}+5 d^{15})-63365 b^{12} (c^{18}+408 c^{15} d^3+9282 c^{12} d^6\\
 &+24310 c^9 d^9+9282 c^6 d^{12}+408 c^3 d^{15}+d^{18})-290 b^3 c^3 d^3 (15 c^{21}\\
 &+1518 c^{18} d^3+24035 c^{15} d^6+89148 c^{12} d^9+89148 c^9 d^{12}+24035 c^6 d^{15}\\
 &+1518 c^3 d^{18}+15 d^{21})-435 b^6 (c^{24}+1012 c^{21} d^3+67298 c^{18} d^6\\
 &+653752 c^{15} d^9+1352078 c^{12} d^{12}+653752 c^9 d^{15}+67298 c^6 d^{18}\\
 &+1012 c^3 d^{21}+d^{24})-435 a^{24} (b^6+c^6-10 c^3 d^3+d^6-10 b^3 (c^3+d^3))\\
 &-63365 a^{18} (b^{12}+c^{12}-110 c^9 d^3+462 c^6 d^6-110 c^3 d^9+d^{12}-110 b^9 (c^3+d^3)\\
 &+462 b^6 (c^6+d^6)-110 b^3 (c^9+d^9))+25852920 a^{15} (b^{12} (c^3-d^3)-11 b^9 (c^6-d^6)\\
 &+11 b^6 (c^9-d^9)+c^3 d^3 (c^9-11 c^6 d^3+11 c^3 d^6-d^9)+b^3 (-c^{12}+d^{12}))\\
 &-63365 a^{12} (b^{18}+c^{18}-408 c^{15} d^3+9282 c^{12} d^6-24310 c^9 d^9+9282 c^6 d^{12}\\
 &-408 c^3 d^{15}+d^{18}-408 b^{15} (c^3+d^3)+9282 b^{12} (c^6+d^6)-24310 b^9 (c^9+d^9)\\
 &+9282 b^6 (c^{12}+d^{12})-408 b^3 (c^{15}+d^{15}))+1394030 a^9 (5 b^{18} (c^3-d^3)-204 b^{15} (c^6-d^6)\\
 &+1105 b^{12} (c^9-d^9)-1105 b^9 (c^{12}-d^{12})+c^3 d^3 (5 c^{15}-204 c^{12} d^3+1105 c^9 d^6\\
 &-1105 c^6 d^9+204 c^3 d^{12}-5 d^{15})+204 b^6 (c^{15}-d^{15})-5 b^3 (c^{18}-d^{18}))\\
 &-435 a^6 (b^{24}+c^{24}-1012 c^{21} d^3+67298 c^{18} d^6-653752 c^{15} d^9\\
 &+1352078 c^{12} d^{12}-653752 c^9 d^{15}+67298 c^6 d^{18}-1012 c^3 d^{21}+d^{24}-1012 b^{21} (c^3+d^3)\\
 &+67298 b^{18} (c^6+d^6)-653752 b^{15} (c^9+d^9)+1352078 b^{12} (c^{12}+d^{12}) \\
 &-653752 b^9 (c^{15}+d^{15})+67298 b^6 (c^{18}+d^{18})-1012 b^3 (c^{21}+d^{21}))\\
 &+290 a^3 (15 b^{24} (c^3-d^3)-1518 b^{21} (c^6-d^6)+24035 b^{18} (c^9-d^9)\\
 &-89148 b^{15} (c^{12}-d^{12})+89148 b^{12} (c^{15}-d^{15})-24035 b^9 (c^{18}-d^{18})\\
 &+c^3 d^3 (15 c^{21}-1518 c^{18} d^3+24035 c^{15} d^6-89148 c^{12} d^9+89148 c^9 d^{12}\\
 &-24035 c^6 d^{15}+1518 c^3 d^{18}-15 d^{21})+1518 b^6 (c^{21}-d^{21})-15 b^3 (c^{24}-d^{24})) .
\end{aligned}
\eeq

\section{Acknowledgements}

One of us (P. L.) would like to acknowledge financial support from the MTA-BME
Kondenz\'alt Anyagok Fizik\'aja Kutat\'ocsoport under grant no: 04119.

\end{document}